\documentclass[%
 reprint,footinbib,
superscriptaddress,
 floatfix,
 amsmath,amssymb,
 aps,
prapp, 
longbibliography,
floatfix,
]{revtex4-2}

\usepackage{graphicx}  
\usepackage{dcolumn}   
\usepackage{bm}        
\usepackage{amssymb}   
\usepackage{amsmath}
\usepackage{color}
\usepackage[colorlinks=true, pdfstartview=FitV, linkcolor=blue, citecolor=blue, urlcolor=blue]{hyperref} 
\usepackage{epstopdf}
\usepackage{siunitx}
\usepackage{tabularx} 
\usepackage{xspace}
\usepackage{braket}
\usepackage{color}
\usepackage{setspace}
\usepackage{mdframed}
\usepackage{framed}
\usepackage{multirow}
\usepackage{setspace}
\usepackage[export]{adjustbox}
\usepackage{varwidth}
\usepackage{float}

\usepackage{lipsum}

\newcommand{\affilETH}{Laboratory for Solid State Physics, ETH Z\"{u}rich, 8093 Z\"urich, Switzerland.}
\newcommand{\affilEMPA}{Laboratory for Magnetic and Functional Thin Films, Empa, Swiss Federal Laboratories for Materials Science and Technology, Dübendorf, 8600 Switzerland.}

\newcommand{\affilUnibas}{Department of Physics, University  of  Basel, 4056  Basel, Switzerland.}

\newcommand{\affilETHQuantum}{Quantum Center, ETH Zurich, 8093 Zurich, Switzerland.}

\begin{document}

\title{Differential Magnetic Force Microscopy with a Switchable Tip}

\author{Shobhna Misra}
\affiliation{\affilETH}

\author{Reshma Peremadathil Pradeep}
\affiliation{\affilEMPA}

\author{Yaoxuan Feng}
\affiliation{\affilEMPA}

\author{Urs Grob}
\affiliation{\affilETH}

\author{Andrada Oana Mandru}
\affiliation{\affilEMPA}

\author{Christian L. Degen}
\affiliation{\affilETH}
\affiliation{\affilETHQuantum}

\author{Hans J. Hug}
\affiliation{\affilEMPA}
\affiliation{\affilUnibas}

\author{Alexander Eichler}
\email[Corresponding author: ]{eichlera@ethz.ch}
\affiliation{\affilETH}
\affiliation{\affilETHQuantum}

\date{\today} 

\begin{abstract}
The separation of physical forces acting on the tip of a magnetic force microscope (MFM) is essential for correct magnetic imaging. Electrostatic forces can be modulated by varying the tip-sample potential and minimized to map the local Kelvin potential. However, distinguishing magnetic forces from van der Waals forces typically requires two measurements with opposite tip magnetizations under otherwise identical measurement conditions. Here, we present an inverted magnetic force microscope where the sample is mounted on a flat cantilever for force sensing, and the magnetic tip is attached to a miniaturized electromagnet that periodically flips the tip’s magnetization. This setup enables the extraction of magnetic tip-sample interactions from the sidebands occurring at the switching rate in the cantilever oscillation spectrum. Our method achieves the separation of magnetic signals from other force contributions in a single-scan mode. Future iterations of this setup may incorporate membrane, trampoline, or string resonators with ultra-high quality factors, potentially improving measurement sensitivity by up to three orders of magnitude compared to the state-of-the-art MFM systems using cantilevers. 
\end{abstract}

\maketitle

\section{Introduction} 
Scanning force microscopy is a versatile method for the characterization and imaging of microscopic objects~\cite{christensen20242024}. Magnetic force detection, in particular, has found widespread use in magnetic force microscopy (MFM) for the imaging of stray magnetic fields emanating from sample surfaces\,\cite{christensen2024}, as well as in magnetic resonance force microscopy (MRFM) for non-invasive, three-dimensional imaging of nanoscale objects~\cite{sidles1991,rugar_2004single,garner2004force,Degen_2009,poggio2010force,vinante2011magnetic,Nichol_2012,Nichol_2013,moores_2015accelerated,haas2022nuclear,Rose_2018,Grob_2019}. In both types of experiments, forces or force gradients between a sharp magnetic tip and sample magnetism are measured as a function of scanner position to create an image. MFM is widely used for the investigation of data storage media\,\cite{Rugar1990,Moser2005,Moser2006}, magnetic thin films\,\cite{Bochi1995,Mandru2020}, pinned uncompensated spins in exchange bias samples\,\cite{Kappenberger2003,Schmid2010}, permanent magnetic materials\,\cite{Neu2018}, superconductors\,\cite{Moser1995,Straver2008,Auslaender2009,Grebenchuk2022}, and current carrying structures\,\cite{Weber2000}, and can also serve to manipulate micromagnetic structures\,\cite{Straver2008,Auslaender2009,Casiraghi2019}. MRFM, on the other hand, has been proposed as a method for understanding the atomic structure of single proteins and other biological molecules, or of nanoscale solid-state devices~\cite{sidles1991,Degen_2009,budakian2024roadmap}.

Generally, a scanning force microscope measures the sum of all forces acting on the tip. Among these are van der Waals and electrostatic forces, which can be of the same order or even larger than the magnetic force\,\cite{Feng2022}. While the electrostatic forces arising from contact potential differences can be locally compensated\,\cite{Feng2022}, differential imaging techniques are required to separate magnetic signals from the van der Waals force originating from local changes of the tip-sample distance\,\cite{Jaafar2011}. This is typically achieved by calculating the differences and sums from two data sets acquired with opposite tip magnetizations, but otherwise identical conditions\,\cite{Feng2022}.

Such `differential scans' can be challenging, particularly for long image acquisition times, or when drift or creep of the piezo electric scanners leads to time-dependent distortions of the acquired images.

Here we report the design and implementation of a magnetic tip that can be switched in situ by current pulses applied via a micro coil. With this device, magnetic and non-magnetic forces can be directly separated within a single passage. Using switching times as short as \SI{40}{\micro\second}, continuous modulation of the tip magnetization with kHz rates is achieved. Our method allows encoding microscopic magnetic information in sidebands of the mechanical cantilever sensor, thereby enabling advanced differential measurement protocols.

\section{Tip Design and Test Setup}

\begin{figure*}[t]
    \centering
    \includegraphics[width=\textwidth]{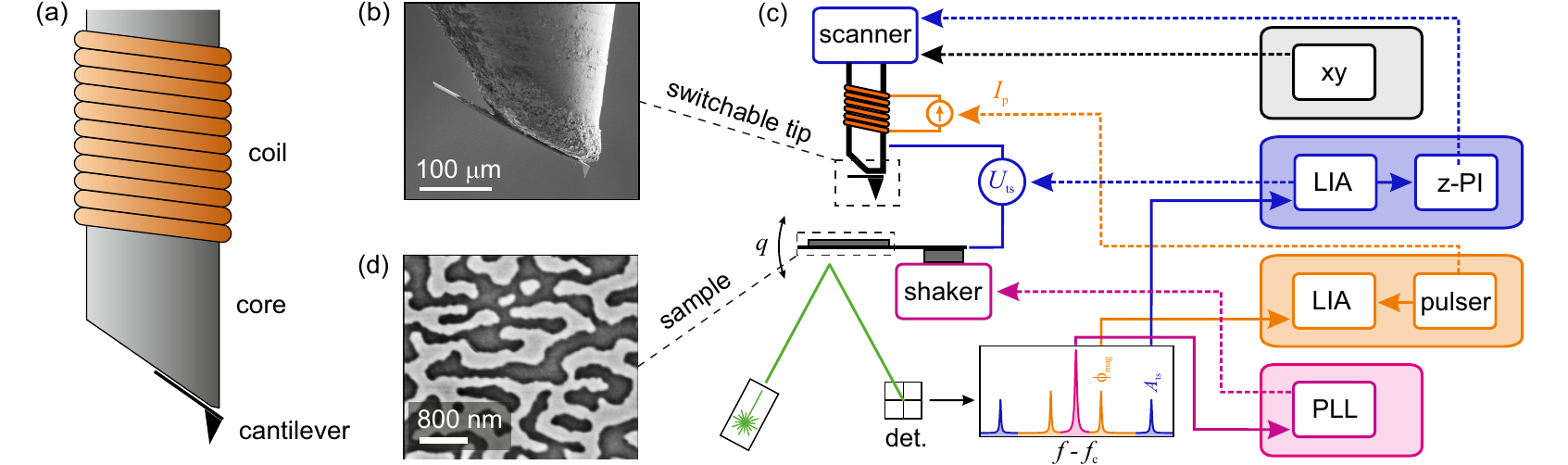}
    \caption{Schematics of the switchable tip and the test setup. (a)~The tip assembly consists of a ferrite core with a coil would around it. The tip is made by coating an AFM probe with magnetic material, breaking it at the base, and glueing it to the tapered end of the core. (b)~Scanning electron micrograph of the AFM probe attached to the bottom of the core. (c)~Control diagram of the inverted MFM setup. The tip assembly is mounted onto the scanner (where the sample would reside in a conventional MFM setup). The magnetic multilayer sample is deposited on a tip-less cantilever, whose position $q$ is detected using optical beam deflection. The cantilever is driven on resonance by a PLL (purple box). For tip-sample distance control, the tip-sample bias is modulated at frequency $f_{\rm ac}$. The amplitudes of the side-bands occurring at $f_{\rm c}\pm 2f_{\rm ac}$ are kept constant by a tip-sample distance feedback (blue box). To periodically switch the tip magnetization, a pulser driving the coil wound around the soft magnetic core is used (orange box). An additional LIA (orange box) is employed to detect the side bands occurring at $f_{\rm c}\pm f_{\rm p}$. (d)~Domain pattern observed on the Co/Pt-multilayer sample deposited onto the cantilever.}
    \label{fig:Fig1}
\end{figure*}

Our design of a switchable magnetic tip is schematically shown in Fig.~\ref{fig:Fig1}(a). We use a Team Nanotec ISC atomic force microscopy (AFM) cantilever, whose tip is coated with a tri-layer of Ta(\SI{2}{\nano\meter})/Co(\SI{5.5}{\nano\meter})/Ta(\SI{4}{\nano\meter}). The tip is glued to the slanted end of a magnetically soft cylindrical core made from MnZn ferrite, with a length of \SI{10}{\milli\meter} and a diameter of \SI{1}{\milli\meter}, see Fig.~\ref{fig:Fig1}(b). The magnetization direction of the core can be controlled by a coil comprising 100 windings (in two layers) of a \SI{50}{\micro\meter} diameter Cu wire. Driven by a current of 
$I =\SI{400}{\milli\ampere}$, a magnetic flux density $B = \SI{60}{\milli\tesla}$ is expected at the tip position, which is sufficient to flip the tip magnetization. See SI for details on numerical simulation.

Our tip-on-magnet setup assembly is mounted onto the sample scanning stage of a home-built MFM operated in vacuum for increased measurement sensitivity~\cite{Feng2022}. Opposite to a conventional MFM setup, the sample, a
Ta(3)/Pt(10)/[Co((0.6)/Pt(1)]$_{5}$/Pt(2) multilayer, is sputter-deposited onto the end of a tip-less Nanosensors TL-FM cantilever ($f_{\rm c} = \SI{75}{\kilo\hertz}$) for sensing the force derivative along the oscillation direction of the cantilever, see Fig.~\ref{fig:Fig1}(c). The tip-less cantilever is driven at its natural resonance frequency $f_{\rm c}$ at a constant amplitude with a phase-locked loop (PLL), see pink box in Fig.~\ref{fig:Fig1}(c). In this configuration, the tip-sample interaction leads to a shift of $f_{\rm c}$, labeled as $df$, governed by the derivative of the sum of magnetic and non-magnetic forces~\cite{Feng2022}.

During scanning, the average tip-sample distance is kept constant. An oscillating tip-sample bias is applied to the cantilever at $f_{\rm ac}=1\,$kHz, leading to sidebands at $f_{\rm c}\pm 2f_{\rm ac}$, see blue box in Fig.~\ref{fig:Fig1}(c). The amplitudes of these sidebands are proportional to the second derivative of the tip-sample capacitance along the oscillatory motion of the cantilever (canted from the z axis by \SI{12}{\degree}). A slow feedback loop (z-PI) adjusts the average tip-sample distance to keep the amplitudes of the sidebands constant. 

For periodic switching of the tip magnetization, a pulser operating at a frequency $f_{\rm p} = 680$\,Hz is connected to the coil, see orange box in Fig.~\ref{fig:Fig1}(c). The periodic switching generates sidebands at $f_{\rm c}\pm f_{\rm p}$ that are measured with the lock-in amplifier (LIA). The absolute amplitude of these sidebands depends on the strength of the magnetic tip-sample interaction force gradient, and their phase is \SI{0}{\degree} or \SI{180}{\degree}, depending on the sign of the magnetic interaction.

\section{Single Pulse Operation}

\begin{figure}
    \centering
    \includegraphics[]{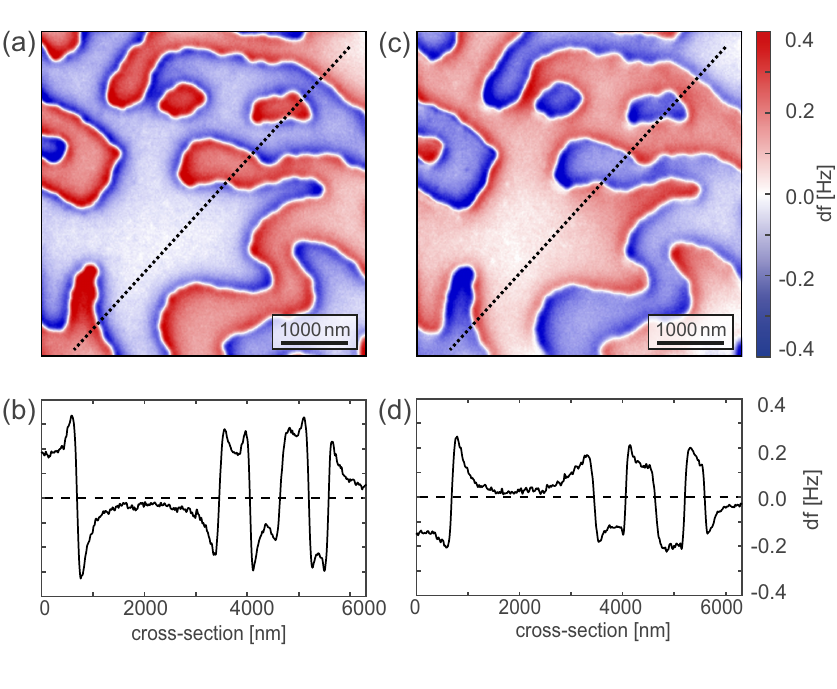}
    \caption{Demonstration of successful tip magnetization switching. (a)~MFM frequency shift data recorded with an `up' tip. See SI for details of the post-processing. (b)~Cross-section along the black line in (a). (c)~MFM frequency shift data recorded with a `down' tip, after a pulse with $t_\mathrm{p} = \SI{100}{\micro\second}$ and $I_\mathrm{p}=\SI{230}{\milli\ampere}$. (d)~Cross-section along the black line in (c).}
    \label{fig:Fig2}
\end{figure}

A first set of experiments is conducted to test the reliability of the tip magnetization switching. Starting with a tip with an unknown magnetization, a single square current pulse of duration $t_\mathrm{p} = \SI{100}{\micro\second}$ and amplitude $I_\mathrm{p}=\SI{230}{\milli\ampere}$ is applied to the coil to set the tip into an `up' magnetization state. Figure~\ref{fig:Fig2}(a) shows the frequency shift recorded with positive tip magnetization. Note that the frequency shift data are processed to remove asymmetries of the domain walls arising from the canted oscillation of the cantilever relative to the surface normal as described in ~\cite{Feng2022}.
Because the tip is in an `up' magnetization state, magnetic domains in the sample with `up' or `down' magnetization lead to an attractive or repulsive tip-sample force, and the corresponding field gradient induces a negative (blue) or positive (red) frequency shift, respectively. The cross-section displayed in Figure~\ref{fig:Fig2}(b), taken along the dashed line in Figure~\ref{fig:Fig2}(a), reveals high contrast near domain walls and low contrast inside larger domains. This is due to the fact that uniform domains with $d/t\rightarrow \infty$ generate negligible stray fields where $d$ is the domain size and $t$ is the magnetic multilayer thickness.  

Next, a pulse with opposite polarity is applied without retracting the tip and a second scan is taken, see Fig.~\ref{fig:Fig2}(c). The data clearly show the same domain pattern as in Fig.~\ref{fig:Fig2}(a) but with opposite magnetic contrast, demonstrating that the tip magnetization can be successfully flipped with a single current pulse. The opposite magnetic contrast is also apparent from a comparison of the cross-sections displayed in Figs.~\ref{fig:Fig2}(b) and (d). The magnitude of the frequency shift is lower in Figs.~\ref{fig:Fig2}(d) than in (b) due to a slightly higher tip-sample distance.

\begin{figure}
    \centering
    \includegraphics[]{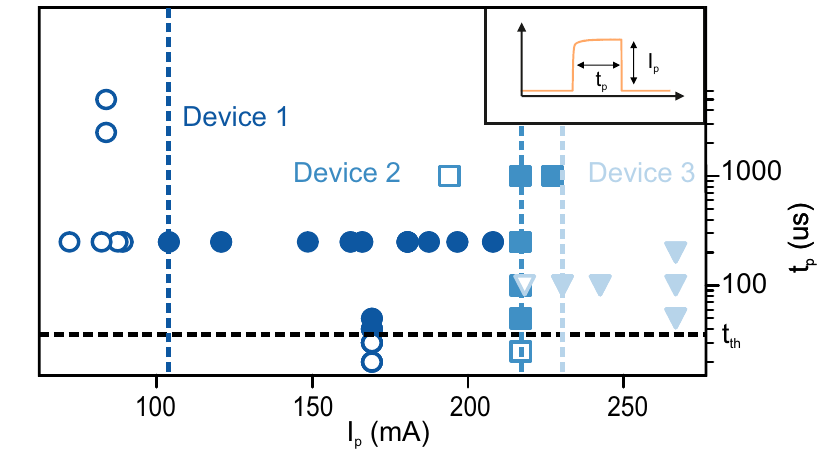}
    \caption{Dependence of the reliability of the tip switching on the applied pulse lengths and currents. Open shapes and filled shapes correspond to failed and successful single-pulse switches, respectively. Different shapes indicate three different tip devices. The inset shows the measured pulse current as a function of time, see SI.}
    \label{fig:Fig3}
\end{figure}

To study the reliability of the tip switching, we perform a series of measurements with varying pulse lengths and current amplitudes. Figure~\ref{fig:Fig3} summarizes the results from three tips. Different shapes (i.e. circles, squares and triangles) show the different devices. Solid shapes indicate that the switching was successful, while hollow shapes indicate that no switching occurred, or switching was unreliable. The magnetization of all tips could be switched reproducibly with pulse lengths down to $t_\mathrm{p}~\approx \SI{40}{\micro\second}$, limited likely by the self-inductance of the coil-core assembly, see SI for details. The minimum current required to switch a tip magnetization varied between $I_\mathrm{p} = \SI{110}{\milli\ampere}$ and $I_\mathrm{p} = \SI{230}{\milli\ampere}$ for the three tips.

\section{Continuous Operation Mode}\label{sec:continuous} 
In a final set of experiments, we demonstrate sideband detection of MFM data obtained by flipping of the tip magnetization continuously with frequency $f_\mathrm{p}$. The flipping creates a time-dependent force gradient that modulates the cantilever frequency at $f_\mathrm{p}$, creating sidebands at $f_\mathrm{c}\pm f_\mathrm{p}$ that we measure with two dedicated LIA demodulators. 
 
In principle, a LIA outputs either the $X$ and $Y$ signal quadratures or the absolute amplitude $R=(X^2+Y^2)^{1/2}$ and the phase $\phi = \tan^{-1}(Y/X)$. When two LIA demodulators are operating at $f_\mathrm{c} \pm f_\mathrm{p}$, $R(f_\mathrm{c}\pm f_\mathrm{p})$ reflects the local magnetic interaction strength, while $\phi(f_\mathrm{c}\pm f_\mathrm{p})$ distinguishes between repulsive or attractive interaction. However, the need to combine amplitude and phase data to obtain magnetic domain images makes this choice inconvenient. Instead, we choose to measure the LIA quadratures $X$ and $Y$. By tuning the reference phase of the LIA, we minimize the $X$ signal, such that the magnetic signal appears only in the $Y$ channel. 

\begin{figure}[t]
    \centering\includegraphics[width=\linewidth]{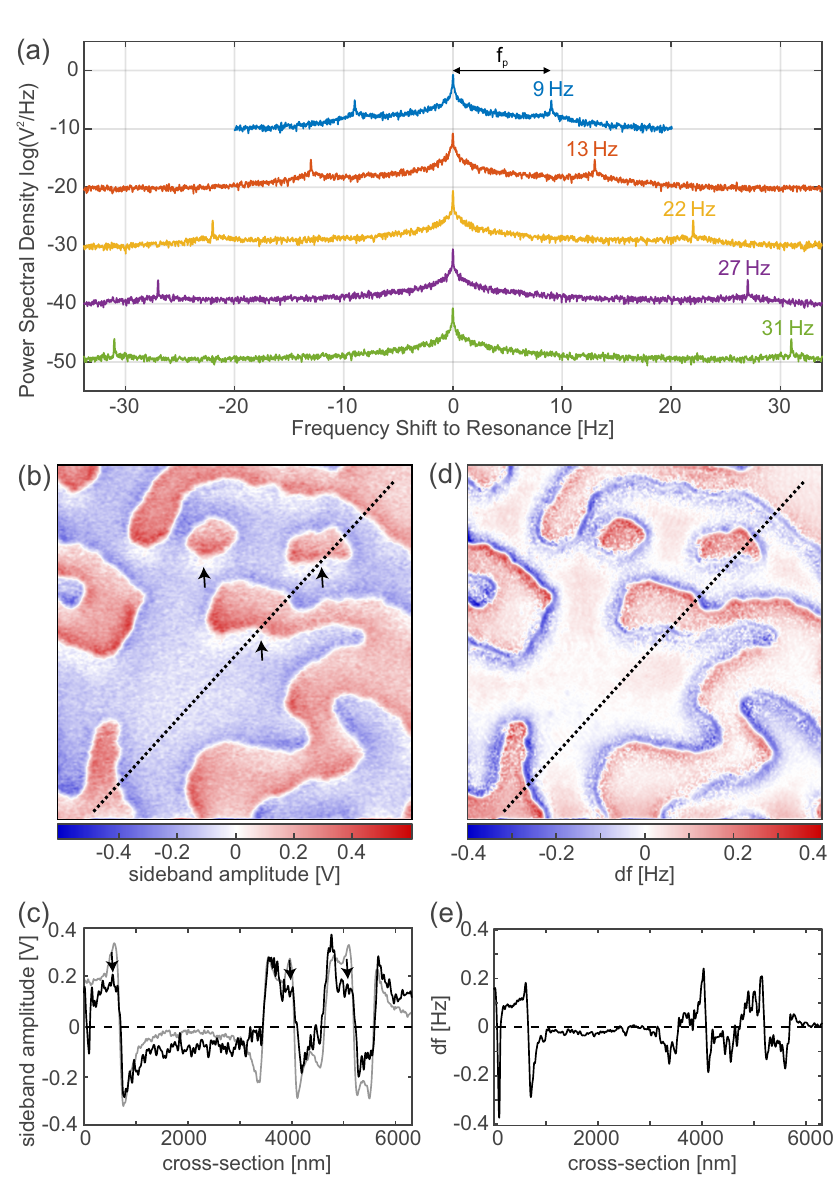}
    \caption{Continuous tip magnetization switching with device 1 and using $I_\mathrm{p} = \SI{230}{\milli\ampere}$. (a)~
    Power spectral density of the cantilever resonance peak at 0\,Hz and the sideband peaks occurring at $f_{\rm c}\pm f_{\rm p}$, with $f_{\rm p} = 9,\,13,\,22,\,27,$ and $31\,$Hz. The spectra of the different frequencies are vertically shifted for clarity. (b)~Sideband image of the magnetic tip-sample interaction, obtained by applying current pulses of duration \SI{100}{\micro\second}, separated by breaks of \SI{1.4}{\milli\second}. Note the distortions at the domain walls (black arrows). (c)~Cross-section (black line) at the location of the dotted line in (b) shown together with a scaled cross-section (faint gray line) taken from Figure~\ref{fig:Fig2}(a). (d)~Frequency shift signal (for details consult the main text). (e)~Cross-section of the frequency shift signal taken at the position of the dotted line in (d). }
    \label{fig:Cont_Operation}
\end{figure}

In Fig.~\ref{fig:Cont_Operation}(a), we plot the power spectral density (PSD) recorded for different $f_{\rm p}$. For this experiment, we use relatively slow modulation $f_{\rm p}$, resulting in clean spectral signatures of the driven cantilever resonance at $f_{\rm c}$ and sidebands at $f_\mathrm{c}\pm f_\mathrm{p}$. For faster modulation, the sidebands become more challenging to distinguish by eye due to the presence of other features in the spectrum (such as the sidebands used for $z$ control). Nevertheless, the sidebands can be tracked with the LIA to extract the magnetic information. 

Fig.~\ref{fig:Cont_Operation}(b) shows an MFM image reconstructed from the $Y$ channel data using $f_{\rm p}=\SI{680}{\hertz}$. Clearly, we find the same domain pattern as in Fig.~\ref{fig:Fig2}(a), albeit with a lower signal-to-noise ratio (SNR). We believe that the lower SNR of the sideband method is due to a higher noise background in the spectrum (against which $f_{\rm p}$ is measured) and due to additional noise in the entire electrical path. These noise sources could be reduced in the future. 

Apart from the additional noise present in the data, we observe a slight distortion of the measured magnetic signal even after correcting for the canted oscillation of the cantilever~\cite{Feng2022}, see black arrows in Fig.~\ref{fig:Cont_Operation}(b). This distortion is also apparent when comparing the cross-sections of Fig.~\ref{fig:Cont_Operation}(c) and Fig.~\ref{fig:Fig2}(b). In parallel to the sideband detection, we also measured the conventional frequency shift signal of the cantilever, see Fig.~\ref{fig:Cont_Operation}(d) and (e). This signal is expected to vanish, as the rapidly switching tip-sample interaction should have a zero mean. However, the measurements reveal a faint domain structure.

We tentatively attribute the distortion in Fig.~\ref{fig:Cont_Operation}(b) and the structure in Fig.~\ref{fig:Cont_Operation}(d) to a net magnetization of the sample, which interacts with the magnetic moment of the ferrite core. The interaction between the ferrite core and the sample magnetization causes the cantilever to bend away from or towards the magnetic tip, depending on the ferrite polarization. Due to this bending, the positive and negative interactions between the tip and local domains in the sample do not cancel perfectly in Fig.~\ref{fig:Cont_Operation}(d), but result in weak average frequency shifts, as observed in Fig.~\ref{fig:Cont_Operation}(d). Because of the canting of the cantilever, the up/down bending also leads to a slight shift of the tip position along the y-direction, which may also explain the weak domain wall distortions visible in Fig.~\ref{fig:Cont_Operation}(b). 

\section{Discussion and Outlook}
The separation of various forces acting simultaneously on a scanning force microscope tip is a challenging experimental task. To date, differential measurements are typically obtained by recording consecutive images~\cite{Feng2022}. This requires tedious calibration to ensure identical measurement conditions between the two images and is often prone to stage drift and piezo creep. In this work, we demonstrate that by using a switchable tip whose magnetization is flipped continuously that magnetic images can be obtained in a single run. In our measurement, the magnetic tip-sample interaction is encoded in a sideband of the cantilever oscillation whose amplitude and phase are unaffected by electrostatic and van der Waals forces. This permits real-time force separation reminiscent to procedures applied for Kelvin probe microscopy, making the measurement less susceptible to drift and other artifacts.  

One factor limiting the switching frequency $f_\mathrm{p}$ in our setup is Joule heating in the coil. Faster switching corresponds to a larger number of pulses, and therefore to a higher duty cycle for a fixed minimum pulse duration. In some cases, we observed that this heating leads to slight distortions in the images, possibly due to thermally induced variations in the tip-sample distance. To avoid such effects, future implementations may employ microfabricated coils placed only a few tens of micrometers from the tip, reminiscent of those used in hard-disk write heads~\cite{Tao2016}. This would not only permit the use of smaller currents and shorter pulses, but also considerably reduce the heat load and the switching field created at the sample location.

One particular application for our switchable tip can be the investigation of magnetic surface defects, which are thought to cause dissipation and decoherence in trapped ions~\cite{turchette_heating_2000, labaziewicz_temperature_2008, safavi-naini_microscopic_2011, brownnutt_ion-trap_2015, kumph_electric-field_2016}, superconducting Josephson circuits~\cite{gao_experimental_2008, wang_improving_2009}, nanomechanical scanning force sensors~\cite{stipe_noncontact_2001, zurita-sanchez_friction_2004, kuehn_dielectric_2006, yazdanian_dielectric_2008, kisiel_suppression_2011, she_noncontact_2012, den_haan_spin-mediated_2015, de_voogd_dissipation_2017,heritier_2021} and nitrogen-vacancy centers in diamond~\cite{PhysRevLett.112.147602}. Understanding which defects are responsible for such effects relies crucially on a probe that can differentiate between magnetic, electrical, and topography contributions. With our setup, this functionality is available by modulating the magnetic moment and the electrical potential of the tip at different frequencies, thereby encoding different information in different sidebands that can be tracked simultaneously.

A second potential area of application is magnetic resonance force microscopy (MRFM)~\cite{sidles1991,Degen_2009}, a variant of nanoscale magnetic resonance imaging~\cite{budakian2024roadmap}. There, switchable magnetic tips might enable the realization of parametric upconversion protocols between nearby mechanical modes of MHz mechanical sensors ~\cite{Kosata_2020,Halg_2022}. In addition, switchable tips offer a simple way for pulsed force modulation to avoid spurious mechanical driving, as implemented in various forms in Refs.~\cite{Nichol_2012,longenecker2012high}.

Our inverted MFM geometry is functionally similar to that used in a previous membrane-based scanning force setup~\cite{Halg_2021}. In a next step, we will thus transfer the demonstrated differential magnetic measurement protocols to membrane or string sensors made from silicon nitride. Performing magnetic force sensing with such resonators instead of silicon cantilevers will ultimately pave the way for more sensitive experiments~\cite{eichler2022ultra}. The most sensitive device demonstrated to date has a room-temperature damping coefficient of $\gamma = \SI{1e-17}{\kilo\gram\per\second}$~\cite{Bereyhi_2022}, corresponding to a single-sided force noise PSD of $S_\mathrm{ff} = \SI{1.6e-37}{\newton\squared\per\hertz}$. For comparison, the best cantilevers used for MFM currently reach $S_\mathrm{ff} = \SI{1.5e-31}{\newton\squared\per\hertz}$ at room temperature. Hence, there is the potential for an improvement of a factor $10^6$ in power sensitivity and $10^3$ in amplitude sensitivity. Such improvement would enormously boost the range of magnetic force experiments that are feasible, including the detection of individual nuclear spins.

\begin{acknowledgments}
We gratefully acknowledge technical support by the mechanical workshop of D-PHYS at ETH Zurich and Sibatron AG. A. E. acknowledges financial support from the Swiss National Science Foundation (SNSF) through grant 200021\_200412 and the ETH Office of Research through grant ETH-51 19-2. 
\end{acknowledgments}

\bibliography{references.bib}

\providecommand{\noopsort}[1]{}\providecommand{\singleletter}[1]{#1}%
\begin{thebibliography}{55}%
\makeatletter
\providecommand \@ifxundefined [1]{%
 \@ifx{#1\undefined}
}%
\providecommand \@ifnum [1]{%
 \ifnum #1\expandafter \@firstoftwo
 \else \expandafter \@secondoftwo
 \fi
}%
\providecommand \@ifx [1]{%
 \ifx #1\expandafter \@firstoftwo
 \else \expandafter \@secondoftwo
 \fi
}%
\providecommand \natexlab [1]{#1}%
\providecommand \enquote  [1]{``#1''}%
\providecommand \bibnamefont  [1]{#1}%
\providecommand \bibfnamefont [1]{#1}%
\providecommand \citenamefont [1]{#1}%
\providecommand \href@noop [0]{\@secondoftwo}%
\providecommand \href [0]{\begingroup \@sanitize@url \@href}%
\providecommand \@href[1]{\@@startlink{#1}\@@href}%
\providecommand \@@href[1]{\endgroup#1\@@endlink}%
\providecommand \@sanitize@url [0]{\catcode `\\12\catcode `\$12\catcode `\&12\catcode `\#12\catcode `\^12\catcode `\_12\catcode `\%12\relax}%
\providecommand \@@startlink[1]{}%
\providecommand \@@endlink[0]{}%
\providecommand \url  [0]{\begingroup\@sanitize@url \@url }%
\providecommand \@url [1]{\endgroup\@href {#1}{\urlprefix }}%
\providecommand \urlprefix  [0]{URL }%
\providecommand \Eprint [0]{\href }%
\providecommand \doibase [0]{https://doi.org/}%
\providecommand \selectlanguage [0]{\@gobble}%
\providecommand \bibinfo  [0]{\@secondoftwo}%
\providecommand \bibfield  [0]{\@secondoftwo}%
\providecommand \translation [1]{[#1]}%
\providecommand \BibitemOpen [0]{}%
\providecommand \bibitemStop [0]{}%
\providecommand \bibitemNoStop [0]{.\EOS\space}%
\providecommand \EOS [0]{\spacefactor3000\relax}%
\providecommand \BibitemShut  [1]{\csname bibitem#1\endcsname}%
\let\auto@bib@innerbib\@empty
\bibitem [{\citenamefont {Christensen}\ \emph {et~al.}(2024{\natexlab{a}})\citenamefont {Christensen}, \citenamefont {Staub}, \citenamefont {Devidas}, \citenamefont {Kalisky}, \citenamefont {Nowack}, \citenamefont {Webb}, \citenamefont {Andersen}, \citenamefont {Huck}, \citenamefont {Broadway}, \citenamefont {Wagner} \emph {et~al.}}]{christensen20242024}%
  \BibitemOpen
  \bibfield  {author} {\bibinfo {author} {\bibfnamefont {D.~V.}\ \bibnamefont {Christensen}}, \bibinfo {author} {\bibfnamefont {U.}~\bibnamefont {Staub}}, \bibinfo {author} {\bibfnamefont {T.}~\bibnamefont {Devidas}}, \bibinfo {author} {\bibfnamefont {B.}~\bibnamefont {Kalisky}}, \bibinfo {author} {\bibfnamefont {K.}~\bibnamefont {Nowack}}, \bibinfo {author} {\bibfnamefont {J.~L.}\ \bibnamefont {Webb}}, \bibinfo {author} {\bibfnamefont {U.~L.}\ \bibnamefont {Andersen}}, \bibinfo {author} {\bibfnamefont {A.}~\bibnamefont {Huck}}, \bibinfo {author} {\bibfnamefont {D.~A.}\ \bibnamefont {Broadway}}, \bibinfo {author} {\bibfnamefont {K.}~\bibnamefont {Wagner}}, \emph {et~al.},\ }\bibfield  {title} {\bibinfo {title} {2024 roadmap on magnetic microscopy techniques and their applications in materials science},\ }\href@noop {} {\bibfield  {journal} {\bibinfo  {journal} {Journal of Physics: Materials}\ } (\bibinfo {year} {2024}{\natexlab{a}})}\BibitemShut {NoStop}%
\bibitem [{\citenamefont {Christensen}\ \emph {et~al.}(2024{\natexlab{b}})\citenamefont {Christensen}, \citenamefont {Staub}, \citenamefont {Devidas}, \citenamefont {Kalisky}, \citenamefont {Nowack}, \citenamefont {Webb}, \citenamefont {Andersen}, \citenamefont {Huck}, \citenamefont {Broadway}, \citenamefont {Wagner}, \citenamefont {Maletinsky}, \citenamefont {Sar}, \citenamefont {Du}, \citenamefont {Yacoby}, \citenamefont {Collomb}, \citenamefont {Bending}, \citenamefont {Oral}, \citenamefont {Hug}, \citenamefont {Mandru}, \citenamefont {Neu}, \citenamefont {Schumacher}, \citenamefont {Sievers}, \citenamefont {Saito}, \citenamefont {Khajetoorians}, \citenamefont {Hauptmann}, \citenamefont {Baumann}, \citenamefont {Eichler}, \citenamefont {Degen}, \citenamefont {McCord}, \citenamefont {Vogel}, \citenamefont {Fiebig}, \citenamefont {Fischer}, \citenamefont {Hierro-Rodriguez}, \citenamefont {Finizio}, \citenamefont {Dhesi}, \citenamefont {Donnelly}, \citenamefont {Büttner}, \citenamefont {Kfir}, \citenamefont
  {Hu}, \citenamefont {Zayko}, \citenamefont {Eisebitt}, \citenamefont {Pfau}, \citenamefont {Frömter}, \citenamefont {Kläui}, \citenamefont {Yasin}, \citenamefont {McMorran}, \citenamefont {Seki}, \citenamefont {Yu}, \citenamefont {Lubk}, \citenamefont {Wolf}, \citenamefont {Pryds}, \citenamefont {Makarov},\ and\ \citenamefont {Poggio}}]{christensen2024}%
  \BibitemOpen
  \bibfield  {author} {\bibinfo {author} {\bibfnamefont {D.~V.}\ \bibnamefont {Christensen}}, \bibinfo {author} {\bibfnamefont {U.}~\bibnamefont {Staub}}, \bibinfo {author} {\bibfnamefont {T.~R.}\ \bibnamefont {Devidas}}, \bibinfo {author} {\bibfnamefont {B.}~\bibnamefont {Kalisky}}, \bibinfo {author} {\bibfnamefont {K.~C.}\ \bibnamefont {Nowack}}, \bibinfo {author} {\bibfnamefont {J.~L.}\ \bibnamefont {Webb}}, \bibinfo {author} {\bibfnamefont {U.~L.}\ \bibnamefont {Andersen}}, \bibinfo {author} {\bibfnamefont {A.}~\bibnamefont {Huck}}, \bibinfo {author} {\bibfnamefont {D.~A.}\ \bibnamefont {Broadway}}, \bibinfo {author} {\bibfnamefont {K.}~\bibnamefont {Wagner}}, \bibinfo {author} {\bibfnamefont {P.}~\bibnamefont {Maletinsky}}, \bibinfo {author} {\bibfnamefont {T.~v.~d.}\ \bibnamefont {Sar}}, \bibinfo {author} {\bibfnamefont {C.~R.}\ \bibnamefont {Du}}, \bibinfo {author} {\bibfnamefont {A.}~\bibnamefont {Yacoby}}, \bibinfo {author} {\bibfnamefont {D.}~\bibnamefont {Collomb}}, \bibinfo {author} {\bibfnamefont
  {S.}~\bibnamefont {Bending}}, \bibinfo {author} {\bibfnamefont {A.}~\bibnamefont {Oral}}, \bibinfo {author} {\bibfnamefont {H.~J.}\ \bibnamefont {Hug}}, \bibinfo {author} {\bibfnamefont {A.-O.}\ \bibnamefont {Mandru}}, \bibinfo {author} {\bibfnamefont {V.}~\bibnamefont {Neu}}, \bibinfo {author} {\bibfnamefont {H.~W.}\ \bibnamefont {Schumacher}}, \bibinfo {author} {\bibfnamefont {S.}~\bibnamefont {Sievers}}, \bibinfo {author} {\bibfnamefont {H.}~\bibnamefont {Saito}}, \bibinfo {author} {\bibfnamefont {A.~A.}\ \bibnamefont {Khajetoorians}}, \bibinfo {author} {\bibfnamefont {N.}~\bibnamefont {Hauptmann}}, \bibinfo {author} {\bibfnamefont {S.}~\bibnamefont {Baumann}}, \bibinfo {author} {\bibfnamefont {A.}~\bibnamefont {Eichler}}, \bibinfo {author} {\bibfnamefont {C.~L.}\ \bibnamefont {Degen}}, \bibinfo {author} {\bibfnamefont {J.}~\bibnamefont {McCord}}, \bibinfo {author} {\bibfnamefont {M.}~\bibnamefont {Vogel}}, \bibinfo {author} {\bibfnamefont {M.}~\bibnamefont {Fiebig}}, \bibinfo {author} {\bibfnamefont
  {P.}~\bibnamefont {Fischer}}, \bibinfo {author} {\bibfnamefont {A.}~\bibnamefont {Hierro-Rodriguez}}, \bibinfo {author} {\bibfnamefont {S.}~\bibnamefont {Finizio}}, \bibinfo {author} {\bibfnamefont {S.~S.}\ \bibnamefont {Dhesi}}, \bibinfo {author} {\bibfnamefont {C.}~\bibnamefont {Donnelly}}, \bibinfo {author} {\bibfnamefont {F.}~\bibnamefont {Büttner}}, \bibinfo {author} {\bibfnamefont {O.}~\bibnamefont {Kfir}}, \bibinfo {author} {\bibfnamefont {W.}~\bibnamefont {Hu}}, \bibinfo {author} {\bibfnamefont {S.}~\bibnamefont {Zayko}}, \bibinfo {author} {\bibfnamefont {S.}~\bibnamefont {Eisebitt}}, \bibinfo {author} {\bibfnamefont {B.}~\bibnamefont {Pfau}}, \bibinfo {author} {\bibfnamefont {R.}~\bibnamefont {Frömter}}, \bibinfo {author} {\bibfnamefont {M.}~\bibnamefont {Kläui}}, \bibinfo {author} {\bibfnamefont {F.~S.}\ \bibnamefont {Yasin}}, \bibinfo {author} {\bibfnamefont {B.~J.}\ \bibnamefont {McMorran}}, \bibinfo {author} {\bibfnamefont {S.}~\bibnamefont {Seki}}, \bibinfo {author} {\bibfnamefont
  {X.}~\bibnamefont {Yu}}, \bibinfo {author} {\bibfnamefont {A.}~\bibnamefont {Lubk}}, \bibinfo {author} {\bibfnamefont {D.}~\bibnamefont {Wolf}}, \bibinfo {author} {\bibfnamefont {N.}~\bibnamefont {Pryds}}, \bibinfo {author} {\bibfnamefont {D.}~\bibnamefont {Makarov}},\ and\ \bibinfo {author} {\bibfnamefont {M.}~\bibnamefont {Poggio}},\ }\bibfield  {title} {\bibinfo {title} {{2024 roadmap on magnetic microscopy techniques and their applications in materials science}},\ }\href {https://doi.org/10.1088/2515-7639/ad31b5} {\bibfield  {journal} {\bibinfo  {journal} {Journal of Physics: Materials}\ }\textbf {\bibinfo {volume} {7}},\ \bibinfo {pages} {032501} (\bibinfo {year} {2024}{\natexlab{b}})}\BibitemShut {NoStop}%
\bibitem [{\citenamefont {Sidles}(1991)}]{sidles1991}%
  \BibitemOpen
  \bibfield  {author} {\bibinfo {author} {\bibfnamefont {J.~A.}\ \bibnamefont {Sidles}},\ }\bibfield  {title} {\bibinfo {title} {Noninductive detection of single‐proton magnetic resonance},\ }\href {https://doi.org/10.1063/1.104757} {\bibfield  {journal} {\bibinfo  {journal} {Applied Physics Letters}\ }\textbf {\bibinfo {volume} {58}},\ \bibinfo {pages} {2854} (\bibinfo {year} {1991})}\BibitemShut {NoStop}%
\bibitem [{\citenamefont {Rugar}\ \emph {et~al.}(2004)\citenamefont {Rugar}, \citenamefont {Budakian}, \citenamefont {Mamin},\ and\ \citenamefont {Chui}}]{rugar_2004single}%
  \BibitemOpen
  \bibfield  {author} {\bibinfo {author} {\bibfnamefont {D.}~\bibnamefont {Rugar}}, \bibinfo {author} {\bibfnamefont {R.}~\bibnamefont {Budakian}}, \bibinfo {author} {\bibfnamefont {H.~J.}\ \bibnamefont {Mamin}},\ and\ \bibinfo {author} {\bibfnamefont {B.~W.}\ \bibnamefont {Chui}},\ }\bibfield  {title} {\bibinfo {title} {Single spin detection by magnetic resonance force microscopy},\ }\href {https://doi.org/10.1038/nature02658} {\bibfield  {journal} {\bibinfo  {journal} {Nature}\ }\textbf {\bibinfo {volume} {430}},\ \bibinfo {pages} {329} (\bibinfo {year} {2004})}\BibitemShut {NoStop}%
\bibitem [{\citenamefont {Garner}\ \emph {et~al.}(2004)\citenamefont {Garner}, \citenamefont {Kuehn}, \citenamefont {Dawlaty}, \citenamefont {Jenkins},\ and\ \citenamefont {Marohn}}]{garner2004force}%
  \BibitemOpen
  \bibfield  {author} {\bibinfo {author} {\bibfnamefont {S.~R.}\ \bibnamefont {Garner}}, \bibinfo {author} {\bibfnamefont {S.}~\bibnamefont {Kuehn}}, \bibinfo {author} {\bibfnamefont {J.~M.}\ \bibnamefont {Dawlaty}}, \bibinfo {author} {\bibfnamefont {N.~E.}\ \bibnamefont {Jenkins}},\ and\ \bibinfo {author} {\bibfnamefont {J.~A.}\ \bibnamefont {Marohn}},\ }\bibfield  {title} {\bibinfo {title} {Force-gradient detected nuclear magnetic resonance},\ }\href@noop {} {\bibfield  {journal} {\bibinfo  {journal} {Applied physics letters}\ }\textbf {\bibinfo {volume} {84}},\ \bibinfo {pages} {5091} (\bibinfo {year} {2004})}\BibitemShut {NoStop}%
\bibitem [{\citenamefont {Degen}\ \emph {et~al.}(2009)\citenamefont {Degen}, \citenamefont {Poggio}, \citenamefont {Mamin}, \citenamefont {Rettner},\ and\ \citenamefont {Rugar}}]{Degen_2009}%
  \BibitemOpen
  \bibfield  {author} {\bibinfo {author} {\bibfnamefont {C.~L.}\ \bibnamefont {Degen}}, \bibinfo {author} {\bibfnamefont {M.}~\bibnamefont {Poggio}}, \bibinfo {author} {\bibfnamefont {H.~J.}\ \bibnamefont {Mamin}}, \bibinfo {author} {\bibfnamefont {C.~T.}\ \bibnamefont {Rettner}},\ and\ \bibinfo {author} {\bibfnamefont {D.}~\bibnamefont {Rugar}},\ }\bibfield  {title} {\bibinfo {title} {Nanoscale magnetic resonance imaging},\ }\href {https://doi.org/10.1073/pnas.0812068106} {\bibfield  {journal} {\bibinfo  {journal} {Proceedings of the National Academy of Sciences}\ }\textbf {\bibinfo {volume} {106}},\ \bibinfo {pages} {1313} (\bibinfo {year} {2009})}\BibitemShut {NoStop}%
\bibitem [{\citenamefont {Poggio}\ and\ \citenamefont {Degen}(2010)}]{poggio2010force}%
  \BibitemOpen
  \bibfield  {author} {\bibinfo {author} {\bibfnamefont {M.}~\bibnamefont {Poggio}}\ and\ \bibinfo {author} {\bibfnamefont {C.~L.}\ \bibnamefont {Degen}},\ }\bibfield  {title} {\bibinfo {title} {Force-detected nuclear magnetic resonance: recent advances and future challenges},\ }\href {https://doi.org/10.1088/0957-4484/21/34/342001} {\bibfield  {journal} {\bibinfo  {journal} {Nanotechnology}\ }\textbf {\bibinfo {volume} {21}},\ \bibinfo {pages} {342001} (\bibinfo {year} {2010})},\ \bibinfo {note} {publisher: IOP Publishing}\BibitemShut {NoStop}%
\bibitem [{\citenamefont {Vinante}\ \emph {et~al.}(2011)\citenamefont {Vinante}, \citenamefont {Wijts}, \citenamefont {Usenko}, \citenamefont {Schinkelshoek},\ and\ \citenamefont {Oosterkamp}}]{vinante2011magnetic}%
  \BibitemOpen
  \bibfield  {author} {\bibinfo {author} {\bibfnamefont {A.}~\bibnamefont {Vinante}}, \bibinfo {author} {\bibfnamefont {G.}~\bibnamefont {Wijts}}, \bibinfo {author} {\bibfnamefont {O.}~\bibnamefont {Usenko}}, \bibinfo {author} {\bibfnamefont {L.}~\bibnamefont {Schinkelshoek}},\ and\ \bibinfo {author} {\bibfnamefont {T.}~\bibnamefont {Oosterkamp}},\ }\bibfield  {title} {\bibinfo {title} {Magnetic resonance force microscopy of paramagnetic electron spins at millikelvin temperatures},\ }\href {https://doi.org/10.1038/ncomms1581} {\bibfield  {journal} {\bibinfo  {journal} {Nature communications}\ }\textbf {\bibinfo {volume} {2}},\ \bibinfo {pages} {572} (\bibinfo {year} {2011})},\ \bibinfo {note} {publisher: Nature Publishing Group UK London}\BibitemShut {NoStop}%
\bibitem [{\citenamefont {Nichol}\ \emph {et~al.}(2012)\citenamefont {Nichol}, \citenamefont {Hemesath}, \citenamefont {Lauhon},\ and\ \citenamefont {Budakian}}]{Nichol_2012}%
  \BibitemOpen
  \bibfield  {author} {\bibinfo {author} {\bibfnamefont {J.~M.}\ \bibnamefont {Nichol}}, \bibinfo {author} {\bibfnamefont {E.~R.}\ \bibnamefont {Hemesath}}, \bibinfo {author} {\bibfnamefont {L.~J.}\ \bibnamefont {Lauhon}},\ and\ \bibinfo {author} {\bibfnamefont {R.}~\bibnamefont {Budakian}},\ }\bibfield  {title} {\bibinfo {title} {Nanomechanical detection of nuclear magnetic resonance using a silicon nanowire oscillator},\ }\href {https://doi.org/10.1103/PhysRevB.85.054414} {\bibfield  {journal} {\bibinfo  {journal} {Physical Review B}\ }\textbf {\bibinfo {volume} {85}},\ \bibinfo {pages} {054414} (\bibinfo {year} {2012})},\ \bibinfo {note} {publisher: American Physical Society}\BibitemShut {NoStop}%
\bibitem [{\citenamefont {Nichol}\ \emph {et~al.}(2013)\citenamefont {Nichol}, \citenamefont {Naibert}, \citenamefont {Hemesath}, \citenamefont {Lauhon},\ and\ \citenamefont {Budakian}}]{Nichol_2013}%
  \BibitemOpen
  \bibfield  {author} {\bibinfo {author} {\bibfnamefont {J.~M.}\ \bibnamefont {Nichol}}, \bibinfo {author} {\bibfnamefont {T.~R.}\ \bibnamefont {Naibert}}, \bibinfo {author} {\bibfnamefont {E.~R.}\ \bibnamefont {Hemesath}}, \bibinfo {author} {\bibfnamefont {L.~J.}\ \bibnamefont {Lauhon}},\ and\ \bibinfo {author} {\bibfnamefont {R.}~\bibnamefont {Budakian}},\ }\bibfield  {title} {\bibinfo {title} {Nanoscale fourier-transform magnetic resonance imaging},\ }\href {https://doi.org/10.1103/PhysRevX.3.031016} {\bibfield  {journal} {\bibinfo  {journal} {Physical Review X}\ }\textbf {\bibinfo {volume} {3}},\ \bibinfo {pages} {031016} (\bibinfo {year} {2013})},\ \bibinfo {note} {number of pages: 7 Publisher: American Physical Society}\BibitemShut {NoStop}%
\bibitem [{\citenamefont {Moores}\ \emph {et~al.}(2015)\citenamefont {Moores}, \citenamefont {Eichler}, \citenamefont {Tao}, \citenamefont {Takahashi}, \citenamefont {Navaretti},\ and\ \citenamefont {Degen}}]{moores_2015accelerated}%
  \BibitemOpen
  \bibfield  {author} {\bibinfo {author} {\bibfnamefont {B.~A.}\ \bibnamefont {Moores}}, \bibinfo {author} {\bibfnamefont {A.}~\bibnamefont {Eichler}}, \bibinfo {author} {\bibfnamefont {Y.}~\bibnamefont {Tao}}, \bibinfo {author} {\bibfnamefont {H.}~\bibnamefont {Takahashi}}, \bibinfo {author} {\bibfnamefont {P.}~\bibnamefont {Navaretti}},\ and\ \bibinfo {author} {\bibfnamefont {C.~L.}\ \bibnamefont {Degen}},\ }\bibfield  {title} {\bibinfo {title} {Accelerated nanoscale magnetic resonance imaging through phase multiplexing},\ }\href {https://doi.org/10.1063/1.4921409} {\bibfield  {journal} {\bibinfo  {journal} {Applied Physics Letters}\ }\textbf {\bibinfo {volume} {106}},\ \bibinfo {pages} {213101} (\bibinfo {year} {2015})}\BibitemShut {NoStop}%
\bibitem [{\citenamefont {Haas}\ \emph {et~al.}(2022)\citenamefont {Haas}, \citenamefont {Tabatabaei}, \citenamefont {Rose}, \citenamefont {Sahafi}, \citenamefont {Piscitelli}, \citenamefont {Jordan}, \citenamefont {Priyadarsi}, \citenamefont {Singh}, \citenamefont {Yager}, \citenamefont {Poole},\ and\ \citenamefont {{others}}}]{haas2022nuclear}%
  \BibitemOpen
  \bibfield  {author} {\bibinfo {author} {\bibfnamefont {H.}~\bibnamefont {Haas}}, \bibinfo {author} {\bibfnamefont {S.}~\bibnamefont {Tabatabaei}}, \bibinfo {author} {\bibfnamefont {W.}~\bibnamefont {Rose}}, \bibinfo {author} {\bibfnamefont {P.}~\bibnamefont {Sahafi}}, \bibinfo {author} {\bibfnamefont {M.}~\bibnamefont {Piscitelli}}, \bibinfo {author} {\bibfnamefont {A.}~\bibnamefont {Jordan}}, \bibinfo {author} {\bibfnamefont {P.}~\bibnamefont {Priyadarsi}}, \bibinfo {author} {\bibfnamefont {N.}~\bibnamefont {Singh}}, \bibinfo {author} {\bibfnamefont {B.}~\bibnamefont {Yager}}, \bibinfo {author} {\bibfnamefont {P.~J.}\ \bibnamefont {Poole}},\ and\ \bibinfo {author} {\bibnamefont {{others}}},\ }\bibfield  {title} {\bibinfo {title} {Nuclear magnetic resonance diffraction with subangstrom precision},\ }\href {https://doi.org/10.1073/pnas.2209213119} {\bibfield  {journal} {\bibinfo  {journal} {Proceedings of the National Academy of Sciences}\ }\textbf {\bibinfo {volume} {119}},\ \bibinfo {pages} {e2209213119}
  (\bibinfo {year} {2022})},\ \bibinfo {note} {publisher: National Acad Sciences}\BibitemShut {NoStop}%
\bibitem [{\citenamefont {Rose}\ \emph {et~al.}(2018)\citenamefont {Rose}, \citenamefont {Haas}, \citenamefont {Chen}, \citenamefont {Jeon}, \citenamefont {Lauhon}, \citenamefont {Cory},\ and\ \citenamefont {Budakian}}]{Rose_2018}%
  \BibitemOpen
  \bibfield  {author} {\bibinfo {author} {\bibfnamefont {W.}~\bibnamefont {Rose}}, \bibinfo {author} {\bibfnamefont {H.}~\bibnamefont {Haas}}, \bibinfo {author} {\bibfnamefont {A.~Q.}\ \bibnamefont {Chen}}, \bibinfo {author} {\bibfnamefont {N.}~\bibnamefont {Jeon}}, \bibinfo {author} {\bibfnamefont {L.~J.}\ \bibnamefont {Lauhon}}, \bibinfo {author} {\bibfnamefont {D.~G.}\ \bibnamefont {Cory}},\ and\ \bibinfo {author} {\bibfnamefont {R.}~\bibnamefont {Budakian}},\ }\bibfield  {title} {\bibinfo {title} {High-resolution nanoscale solid-state nuclear magnetic resonance spectroscopy},\ }\href {https://doi.org/10.1103/PhysRevX.8.011030} {\bibfield  {journal} {\bibinfo  {journal} {Phys. Rev. X}\ }\textbf {\bibinfo {volume} {8}},\ \bibinfo {pages} {011030} (\bibinfo {year} {2018})}\BibitemShut {NoStop}%
\bibitem [{\citenamefont {Grob}\ \emph {et~al.}(2019)\citenamefont {Grob}, \citenamefont {Krass}, \citenamefont {H\'eritier}, \citenamefont {Pachlatko}, \citenamefont {Rhensius}, \citenamefont {Ko\v{s}ata}, \citenamefont {Moores}, \citenamefont {Takahashi}, \citenamefont {Eichler},\ and\ \citenamefont {Degen}}]{Grob_2019}%
  \BibitemOpen
  \bibfield  {author} {\bibinfo {author} {\bibfnamefont {U.}~\bibnamefont {Grob}}, \bibinfo {author} {\bibfnamefont {M.-D.}\ \bibnamefont {Krass}}, \bibinfo {author} {\bibfnamefont {M.}~\bibnamefont {H\'eritier}}, \bibinfo {author} {\bibfnamefont {R.}~\bibnamefont {Pachlatko}}, \bibinfo {author} {\bibfnamefont {J.}~\bibnamefont {Rhensius}}, \bibinfo {author} {\bibfnamefont {J.}~\bibnamefont {Ko\v{s}ata}}, \bibinfo {author} {\bibfnamefont {B.~A.~J.}\ \bibnamefont {Moores}}, \bibinfo {author} {\bibfnamefont {H.}~\bibnamefont {Takahashi}}, \bibinfo {author} {\bibfnamefont {A.}~\bibnamefont {Eichler}},\ and\ \bibinfo {author} {\bibfnamefont {C.~L.}\ \bibnamefont {Degen}},\ }\bibfield  {title} {\bibinfo {title} {Magnetic resonance force microscopy with a one-dimensional resolution of 0.9 nanometers},\ }\href {https://doi.org/10.1021/acs.nanolett.9b03048} {\bibfield  {journal} {\bibinfo  {journal} {Nano Lett.}\ }\textbf {\bibinfo {volume} {19}},\ \bibinfo {pages} {7935} (\bibinfo {year} {2019})}\BibitemShut
  {NoStop}%
\bibitem [{\citenamefont {Rugar}\ \emph {et~al.}(1990)\citenamefont {Rugar}, \citenamefont {Mamin}, \citenamefont {Guethner}, \citenamefont {Lambert}, \citenamefont {Stern}, \citenamefont {McFadyen},\ and\ \citenamefont {Yogi}}]{Rugar1990}%
  \BibitemOpen
  \bibfield  {author} {\bibinfo {author} {\bibfnamefont {D.}~\bibnamefont {Rugar}}, \bibinfo {author} {\bibfnamefont {H.~J.}\ \bibnamefont {Mamin}}, \bibinfo {author} {\bibfnamefont {P.}~\bibnamefont {Guethner}}, \bibinfo {author} {\bibfnamefont {S.~E.}\ \bibnamefont {Lambert}}, \bibinfo {author} {\bibfnamefont {J.~E.}\ \bibnamefont {Stern}}, \bibinfo {author} {\bibfnamefont {I.}~\bibnamefont {McFadyen}},\ and\ \bibinfo {author} {\bibfnamefont {T.}~\bibnamefont {Yogi}},\ }\bibfield  {title} {\bibinfo {title} {{Magnetic force microscopy: General principles and application to longitudinal recording media}},\ }\href {https://doi.org/10.1063/1.346713} {\bibfield  {journal} {\bibinfo  {journal} {Journal Of Applied Physics}\ }\textbf {\bibinfo {volume} {68}},\ \bibinfo {pages} {1169 } (\bibinfo {year} {1990})}\BibitemShut {NoStop}%
\bibitem [{\citenamefont {Moser}\ \emph {et~al.}(2005)\citenamefont {Moser}, \citenamefont {Xiao}, \citenamefont {Kappenberger}, \citenamefont {Takano}, \citenamefont {Weresin}, \citenamefont {Ikeda}, \citenamefont {Do},\ and\ \citenamefont {Hug}}]{Moser2005}%
  \BibitemOpen
  \bibfield  {author} {\bibinfo {author} {\bibfnamefont {A.}~\bibnamefont {Moser}}, \bibinfo {author} {\bibfnamefont {M.}~\bibnamefont {Xiao}}, \bibinfo {author} {\bibfnamefont {P.}~\bibnamefont {Kappenberger}}, \bibinfo {author} {\bibfnamefont {K.}~\bibnamefont {Takano}}, \bibinfo {author} {\bibfnamefont {W.}~\bibnamefont {Weresin}}, \bibinfo {author} {\bibfnamefont {Y.}~\bibnamefont {Ikeda}}, \bibinfo {author} {\bibfnamefont {H.}~\bibnamefont {Do}},\ and\ \bibinfo {author} {\bibfnamefont {H.}~\bibnamefont {Hug}},\ }\bibfield  {title} {\bibinfo {title} {{High-resolution magnetic force microscopy study of high-density transitions in perpendicular recording media}},\ }\href {https://doi.org/10.1016/j.jmmm.2004.10.048} {\bibfield  {journal} {\bibinfo  {journal} {Journal of Magnetism and Magnetic Materials}\ }\textbf {\bibinfo {volume} {287}},\ \bibinfo {pages} {298} (\bibinfo {year} {2005})},\ \bibinfo {note} {7th Perpendicular Magnetic Recording Conference (PMRC 2004), Sendai, JAPAN, MAY 31-JUN 02,
  2004}\BibitemShut {NoStop}%
\bibitem [{\citenamefont {Moser}\ \emph {et~al.}(2006)\citenamefont {Moser}, \citenamefont {Bonhote}, \citenamefont {Dai}, \citenamefont {Do}, \citenamefont {Knigge}, \citenamefont {Ikeda}, \citenamefont {Le}, \citenamefont {Lengsfield}, \citenamefont {MacDonald}, \citenamefont {Li}, \citenamefont {Nayak}, \citenamefont {Payne}, \citenamefont {Schabes}, \citenamefont {Smith}, \citenamefont {Takano}, \citenamefont {Tsang}, \citenamefont {Heijden}, \citenamefont {Weresin}, \citenamefont {Williams},\ and\ \citenamefont {Xiao}}]{Moser2006}%
  \BibitemOpen
  \bibfield  {author} {\bibinfo {author} {\bibfnamefont {A.}~\bibnamefont {Moser}}, \bibinfo {author} {\bibfnamefont {C.}~\bibnamefont {Bonhote}}, \bibinfo {author} {\bibfnamefont {Q.}~\bibnamefont {Dai}}, \bibinfo {author} {\bibfnamefont {H.}~\bibnamefont {Do}}, \bibinfo {author} {\bibfnamefont {B.}~\bibnamefont {Knigge}}, \bibinfo {author} {\bibfnamefont {Y.}~\bibnamefont {Ikeda}}, \bibinfo {author} {\bibfnamefont {Q.}~\bibnamefont {Le}}, \bibinfo {author} {\bibfnamefont {B.}~\bibnamefont {Lengsfield}}, \bibinfo {author} {\bibfnamefont {S.}~\bibnamefont {MacDonald}}, \bibinfo {author} {\bibfnamefont {J.}~\bibnamefont {Li}}, \bibinfo {author} {\bibfnamefont {V.}~\bibnamefont {Nayak}}, \bibinfo {author} {\bibfnamefont {R.}~\bibnamefont {Payne}}, \bibinfo {author} {\bibfnamefont {M.}~\bibnamefont {Schabes}}, \bibinfo {author} {\bibfnamefont {N.}~\bibnamefont {Smith}}, \bibinfo {author} {\bibfnamefont {K.}~\bibnamefont {Takano}}, \bibinfo {author} {\bibfnamefont {C.}~\bibnamefont {Tsang}}, \bibinfo {author}
  {\bibfnamefont {P.~v.~d.}\ \bibnamefont {Heijden}}, \bibinfo {author} {\bibfnamefont {W.}~\bibnamefont {Weresin}}, \bibinfo {author} {\bibfnamefont {M.}~\bibnamefont {Williams}},\ and\ \bibinfo {author} {\bibfnamefont {M.}~\bibnamefont {Xiao}},\ }\bibfield  {title} {\bibinfo {title} {{Perpendicular magnetic recording technology at 230 Gbit/in(2)}},\ }\href {https://doi.org/10.1016/j.jmmm.2006.01.033} {\bibfield  {journal} {\bibinfo  {journal} {Journal Of Magnetism And Magnetic Materials}\ }\textbf {\bibinfo {volume} {303}},\ \bibinfo {pages} {271 } (\bibinfo {year} {2006})}\BibitemShut {NoStop}%
\bibitem [{\citenamefont {Bochi}\ \emph {et~al.}(1995)\citenamefont {Bochi}, \citenamefont {Hug}, \citenamefont {Paul}, \citenamefont {Stiefel}, \citenamefont {Moser}, \citenamefont {Parashikov}, \citenamefont {Güntherodt},\ and\ \citenamefont {O'Handley}}]{Bochi1995}%
  \BibitemOpen
  \bibfield  {author} {\bibinfo {author} {\bibfnamefont {G.}~\bibnamefont {Bochi}}, \bibinfo {author} {\bibfnamefont {H.~J.}\ \bibnamefont {Hug}}, \bibinfo {author} {\bibfnamefont {D.~I.}\ \bibnamefont {Paul}}, \bibinfo {author} {\bibfnamefont {B.}~\bibnamefont {Stiefel}}, \bibinfo {author} {\bibfnamefont {A.}~\bibnamefont {Moser}}, \bibinfo {author} {\bibfnamefont {I.}~\bibnamefont {Parashikov}}, \bibinfo {author} {\bibfnamefont {H.-J.}\ \bibnamefont {Güntherodt}},\ and\ \bibinfo {author} {\bibfnamefont {R.~C.}\ \bibnamefont {O'Handley}},\ }\bibfield  {title} {\bibinfo {title} {{Magnetic Domain Structure in Ultrathin Films}},\ }\href {https://doi.org/10.1103/physrevlett.75.1839} {\bibfield  {journal} {\bibinfo  {journal} {Physical Review Letters}\ }\textbf {\bibinfo {volume} {75}},\ \bibinfo {pages} {1839} (\bibinfo {year} {1995})}\BibitemShut {NoStop}%
\bibitem [{\citenamefont {Mandru}\ \emph {et~al.}(2020)\citenamefont {Mandru}, \citenamefont {Yıldırım}, \citenamefont {Tomasello}, \citenamefont {Heistracher}, \citenamefont {Penedo}, \citenamefont {Giordano}, \citenamefont {Suess}, \citenamefont {Finocchio},\ and\ \citenamefont {Hug}}]{Mandru2020}%
  \BibitemOpen
  \bibfield  {author} {\bibinfo {author} {\bibfnamefont {A.~O.}\ \bibnamefont {Mandru}}, \bibinfo {author} {\bibfnamefont {O.}~\bibnamefont {Yıldırım}}, \bibinfo {author} {\bibfnamefont {R.}~\bibnamefont {Tomasello}}, \bibinfo {author} {\bibfnamefont {P.}~\bibnamefont {Heistracher}}, \bibinfo {author} {\bibfnamefont {M.}~\bibnamefont {Penedo}}, \bibinfo {author} {\bibfnamefont {A.}~\bibnamefont {Giordano}}, \bibinfo {author} {\bibfnamefont {D.}~\bibnamefont {Suess}}, \bibinfo {author} {\bibfnamefont {G.}~\bibnamefont {Finocchio}},\ and\ \bibinfo {author} {\bibfnamefont {H.-J.}\ \bibnamefont {Hug}},\ }\bibfield  {title} {\bibinfo {title} {{Coexistence of distinct skyrmion phases observed in hybrid ferromagnetic/ferrimagnetic multilayers.}},\ }\href {https://doi.org/10.1038/s41467-020-20025-2} {\bibfield  {journal} {\bibinfo  {journal} {Nature Communications}\ }\textbf {\bibinfo {volume} {11}},\ \bibinfo {pages} {6365 } (\bibinfo {year} {2020})}\BibitemShut {NoStop}%
\bibitem [{\citenamefont {Kappenberger}\ \emph {et~al.}(2003)\citenamefont {Kappenberger}, \citenamefont {Martin}, \citenamefont {Pellmont}, \citenamefont {Hug}, \citenamefont {Kortright}, \citenamefont {Hellwig},\ and\ \citenamefont {Fullerton}}]{Kappenberger2003}%
  \BibitemOpen
  \bibfield  {author} {\bibinfo {author} {\bibfnamefont {P.}~\bibnamefont {Kappenberger}}, \bibinfo {author} {\bibfnamefont {S.}~\bibnamefont {Martin}}, \bibinfo {author} {\bibfnamefont {Y.}~\bibnamefont {Pellmont}}, \bibinfo {author} {\bibfnamefont {H.~J.}\ \bibnamefont {Hug}}, \bibinfo {author} {\bibfnamefont {J.~B.}\ \bibnamefont {Kortright}}, \bibinfo {author} {\bibfnamefont {O.}~\bibnamefont {Hellwig}},\ and\ \bibinfo {author} {\bibfnamefont {E.~E.}\ \bibnamefont {Fullerton}},\ }\bibfield  {title} {\bibinfo {title} {{Direct imaging and determination of the uncompensated spin density in exchange-biased CoO/(CoPt) multilayers.}},\ }\href {https://doi.org/10.1103/physrevlett.91.267202} {\bibfield  {journal} {\bibinfo  {journal} {Physical Review Letters}\ }\textbf {\bibinfo {volume} {91}},\ \bibinfo {pages} {267202} (\bibinfo {year} {2003})}\BibitemShut {NoStop}%
\bibitem [{\citenamefont {Schmid}\ \emph {et~al.}(2010)\citenamefont {Schmid}, \citenamefont {Marioni}, \citenamefont {Kappenberger}, \citenamefont {Romer}, \citenamefont {Parlinska-Wojtan}, \citenamefont {Hug}, \citenamefont {Hellwig}, \citenamefont {Carey},\ and\ \citenamefont {Fullerton}}]{Schmid2010}%
  \BibitemOpen
  \bibfield  {author} {\bibinfo {author} {\bibfnamefont {I.}~\bibnamefont {Schmid}}, \bibinfo {author} {\bibfnamefont {M.~A.}\ \bibnamefont {Marioni}}, \bibinfo {author} {\bibfnamefont {P.}~\bibnamefont {Kappenberger}}, \bibinfo {author} {\bibfnamefont {S.}~\bibnamefont {Romer}}, \bibinfo {author} {\bibfnamefont {M.}~\bibnamefont {Parlinska-Wojtan}}, \bibinfo {author} {\bibfnamefont {H.~J.}\ \bibnamefont {Hug}}, \bibinfo {author} {\bibfnamefont {O.}~\bibnamefont {Hellwig}}, \bibinfo {author} {\bibfnamefont {M.~J.}\ \bibnamefont {Carey}},\ and\ \bibinfo {author} {\bibfnamefont {E.~E.}\ \bibnamefont {Fullerton}},\ }\bibfield  {title} {\bibinfo {title} {{Exchange bias and domain evolution at 10 nm scales}},\ }\href {https://doi.org/10.1103/physrevlett.105.197201} {\bibfield  {journal} {\bibinfo  {journal} {Physical Review Letters}\ }\textbf {\bibinfo {volume} {105}},\ \bibinfo {pages} {197201} (\bibinfo {year} {2010})}\BibitemShut {NoStop}%
\bibitem [{\citenamefont {Neu}\ \emph {et~al.}(2018)\citenamefont {Neu}, \citenamefont {Vock}, \citenamefont {Sturm},\ and\ \citenamefont {Schultz}}]{Neu2018}%
  \BibitemOpen
  \bibfield  {author} {\bibinfo {author} {\bibfnamefont {V.}~\bibnamefont {Neu}}, \bibinfo {author} {\bibfnamefont {S.}~\bibnamefont {Vock}}, \bibinfo {author} {\bibfnamefont {T.}~\bibnamefont {Sturm}},\ and\ \bibinfo {author} {\bibfnamefont {L.}~\bibnamefont {Schultz}},\ }\bibfield  {title} {\bibinfo {title} {{Epitaxial hard magnetic SmCo 5MFM tips – a new approach to advanced magnetic force microscopy imaging}},\ }\href {https://doi.org/10.1039/c8nr03997f} {\bibfield  {journal} {\bibinfo  {journal} {Nanoscale}\ }\textbf {\bibinfo {volume} {10}},\ \bibinfo {pages} {16881 } (\bibinfo {year} {2018})}\BibitemShut {NoStop}%
\bibitem [{\citenamefont {Moser}\ \emph {et~al.}(1995)\citenamefont {Moser}, \citenamefont {Hug}, \citenamefont {Parashikov}, \citenamefont {Stiefel}, \citenamefont {Fritz}, \citenamefont {Thomas}, \citenamefont {Baratoff}, \citenamefont {Güntherodt},\ and\ \citenamefont {Chaudhari}}]{Moser1995}%
  \BibitemOpen
  \bibfield  {author} {\bibinfo {author} {\bibfnamefont {A.}~\bibnamefont {Moser}}, \bibinfo {author} {\bibfnamefont {H.~J.}\ \bibnamefont {Hug}}, \bibinfo {author} {\bibfnamefont {I.}~\bibnamefont {Parashikov}}, \bibinfo {author} {\bibfnamefont {B.}~\bibnamefont {Stiefel}}, \bibinfo {author} {\bibfnamefont {O.}~\bibnamefont {Fritz}}, \bibinfo {author} {\bibfnamefont {H.}~\bibnamefont {Thomas}}, \bibinfo {author} {\bibfnamefont {A.}~\bibnamefont {Baratoff}}, \bibinfo {author} {\bibfnamefont {H.-J.}\ \bibnamefont {Güntherodt}},\ and\ \bibinfo {author} {\bibfnamefont {P.}~\bibnamefont {Chaudhari}},\ }\bibfield  {title} {\bibinfo {title} {{Observation of Single Vortices Condensed into a Vortex-Glass Phase by Magnetic Force Microscopy}},\ }\href {https://doi.org/10.1103/physrevlett.74.1847} {\bibfield  {journal} {\bibinfo  {journal} {Physical Review Letters}\ }\textbf {\bibinfo {volume} {74}},\ \bibinfo {pages} {1847} (\bibinfo {year} {1995})}\BibitemShut {NoStop}%
\bibitem [{\citenamefont {Straver}\ \emph {et~al.}(2008)\citenamefont {Straver}, \citenamefont {Hoffman}, \citenamefont {Auslaender}, \citenamefont {Rugar},\ and\ \citenamefont {Moler}}]{Straver2008}%
  \BibitemOpen
  \bibfield  {author} {\bibinfo {author} {\bibfnamefont {E.~W.~J.}\ \bibnamefont {Straver}}, \bibinfo {author} {\bibfnamefont {J.~E.}\ \bibnamefont {Hoffman}}, \bibinfo {author} {\bibfnamefont {O.~M.}\ \bibnamefont {Auslaender}}, \bibinfo {author} {\bibfnamefont {D.}~\bibnamefont {Rugar}},\ and\ \bibinfo {author} {\bibfnamefont {K.~A.}\ \bibnamefont {Moler}},\ }\bibfield  {title} {\bibinfo {title} {{Controlled manipulation of individual vortices in a superconductor}},\ }\bibfield  {journal} {\bibinfo  {journal} {Applied Physics Letters}\ }\textbf {\bibinfo {volume} {93}},\ \href {https://doi.org/10.1063/1.3000963} {10.1063/1.3000963} (\bibinfo {year} {2008})\BibitemShut {NoStop}%
\bibitem [{\citenamefont {Auslaender}\ \emph {et~al.}(2009)\citenamefont {Auslaender}, \citenamefont {Luan}, \citenamefont {Straver}, \citenamefont {Hoffman}, \citenamefont {Koshnick}, \citenamefont {Zeldov}, \citenamefont {Bonn}, \citenamefont {Liang}, \citenamefont {Hardy},\ and\ \citenamefont {Moler}}]{Auslaender2009}%
  \BibitemOpen
  \bibfield  {author} {\bibinfo {author} {\bibfnamefont {O.~M.}\ \bibnamefont {Auslaender}}, \bibinfo {author} {\bibfnamefont {L.}~\bibnamefont {Luan}}, \bibinfo {author} {\bibfnamefont {E.~W.~J.}\ \bibnamefont {Straver}}, \bibinfo {author} {\bibfnamefont {J.~E.}\ \bibnamefont {Hoffman}}, \bibinfo {author} {\bibfnamefont {N.~C.}\ \bibnamefont {Koshnick}}, \bibinfo {author} {\bibfnamefont {E.}~\bibnamefont {Zeldov}}, \bibinfo {author} {\bibfnamefont {D.~A.}\ \bibnamefont {Bonn}}, \bibinfo {author} {\bibfnamefont {R.}~\bibnamefont {Liang}}, \bibinfo {author} {\bibfnamefont {W.~N.}\ \bibnamefont {Hardy}},\ and\ \bibinfo {author} {\bibfnamefont {K.~A.}\ \bibnamefont {Moler}},\ }\bibfield  {title} {\bibinfo {title} {{Mechanics of individual isolated vortices in a cuprate superconductor}},\ }\href {https://doi.org/10.1038/nphys1127} {\bibfield  {journal} {\bibinfo  {journal} {Nature Physics}\ }\textbf {\bibinfo {volume} {5}},\ \bibinfo {pages} {35} (\bibinfo {year} {2009})},\ \Eprint
  {https://arxiv.org/abs/0809.2817} {0809.2817} \BibitemShut {NoStop}%
\bibitem [{\citenamefont {Grebenchuk}\ \emph {et~al.}(2022)\citenamefont {Grebenchuk}, \citenamefont {Hovhannisyan}, \citenamefont {Shishkin}, \citenamefont {Dremov},\ and\ \citenamefont {Stolyarov}}]{Grebenchuk2022}%
  \BibitemOpen
  \bibfield  {author} {\bibinfo {author} {\bibfnamefont {S.}~\bibnamefont {Grebenchuk}}, \bibinfo {author} {\bibfnamefont {R.}~\bibnamefont {Hovhannisyan}}, \bibinfo {author} {\bibfnamefont {A.}~\bibnamefont {Shishkin}}, \bibinfo {author} {\bibfnamefont {V.}~\bibnamefont {Dremov}},\ and\ \bibinfo {author} {\bibfnamefont {V.}~\bibnamefont {Stolyarov}},\ }\bibfield  {title} {\bibinfo {title} {{Magnetic Force Microscopy for Diagnosis of Complex Superconducting Circuits}},\ }\href {https://doi.org/10.1103/physrevapplied.18.054035} {\bibfield  {journal} {\bibinfo  {journal} {Physical Review Applied}\ }\textbf {\bibinfo {volume} {18}},\ \bibinfo {pages} {054035} (\bibinfo {year} {2022})}\BibitemShut {NoStop}%
\bibitem [{\citenamefont {Weber}\ \emph {et~al.}(2000)\citenamefont {Weber}, \citenamefont {Mertin},\ and\ \citenamefont {Kubalek}}]{Weber2000}%
  \BibitemOpen
  \bibfield  {author} {\bibinfo {author} {\bibfnamefont {R.}~\bibnamefont {Weber}}, \bibinfo {author} {\bibfnamefont {W.}~\bibnamefont {Mertin}},\ and\ \bibinfo {author} {\bibfnamefont {E.}~\bibnamefont {Kubalek}},\ }\bibfield  {title} {\bibinfo {title} {{Voltage-influence of biased interconnection line on integrated circuit-internal current contrast measurements via magnetic force microscopy}},\ }\href {https://doi.org/10.1016/s0026-2714(00)00158-x} {\bibfield  {journal} {\bibinfo  {journal} {Microelectronics Reliability}\ }\textbf {\bibinfo {volume} {40}},\ \bibinfo {pages} {1389} (\bibinfo {year} {2000})}\BibitemShut {NoStop}%
\bibitem [{\citenamefont {Casiraghi}\ \emph {et~al.}(2019)\citenamefont {Casiraghi}, \citenamefont {Corte-León}, \citenamefont {Vafaee}, \citenamefont {Garcia-Sanchez}, \citenamefont {Durin}, \citenamefont {Pasquale}, \citenamefont {Jakob}, \citenamefont {Kläui},\ and\ \citenamefont {Kazakova}}]{Casiraghi2019}%
  \BibitemOpen
  \bibfield  {author} {\bibinfo {author} {\bibfnamefont {A.}~\bibnamefont {Casiraghi}}, \bibinfo {author} {\bibfnamefont {H.}~\bibnamefont {Corte-León}}, \bibinfo {author} {\bibfnamefont {M.}~\bibnamefont {Vafaee}}, \bibinfo {author} {\bibfnamefont {F.}~\bibnamefont {Garcia-Sanchez}}, \bibinfo {author} {\bibfnamefont {G.}~\bibnamefont {Durin}}, \bibinfo {author} {\bibfnamefont {M.}~\bibnamefont {Pasquale}}, \bibinfo {author} {\bibfnamefont {G.}~\bibnamefont {Jakob}}, \bibinfo {author} {\bibfnamefont {M.}~\bibnamefont {Kläui}},\ and\ \bibinfo {author} {\bibfnamefont {O.}~\bibnamefont {Kazakova}},\ }\bibfield  {title} {\bibinfo {title} {{Individual skyrmion manipulation by local magnetic field gradients}},\ }\href {https://doi.org/10.1038/s42005-019-0242-5} {\bibfield  {journal} {\bibinfo  {journal} {Communications Physics}\ }\textbf {\bibinfo {volume} {2}},\ \bibinfo {pages} {1 } (\bibinfo {year} {2019})}\BibitemShut {NoStop}%
\bibitem [{\citenamefont {Budakian}\ \emph {et~al.}(2024)\citenamefont {Budakian}, \citenamefont {Finkler}, \citenamefont {Eichler}, \citenamefont {Poggio}, \citenamefont {Degen}, \citenamefont {Tabatabaei}, \citenamefont {Lee}, \citenamefont {Hammel}, \citenamefont {Polzik}, \citenamefont {Taminiau} \emph {et~al.}}]{budakian2024roadmap}%
  \BibitemOpen
  \bibfield  {author} {\bibinfo {author} {\bibfnamefont {R.}~\bibnamefont {Budakian}}, \bibinfo {author} {\bibfnamefont {A.}~\bibnamefont {Finkler}}, \bibinfo {author} {\bibfnamefont {A.}~\bibnamefont {Eichler}}, \bibinfo {author} {\bibfnamefont {M.}~\bibnamefont {Poggio}}, \bibinfo {author} {\bibfnamefont {C.~L.}\ \bibnamefont {Degen}}, \bibinfo {author} {\bibfnamefont {S.}~\bibnamefont {Tabatabaei}}, \bibinfo {author} {\bibfnamefont {I.}~\bibnamefont {Lee}}, \bibinfo {author} {\bibfnamefont {P.~C.}\ \bibnamefont {Hammel}}, \bibinfo {author} {\bibfnamefont {E.}~\bibnamefont {Polzik}}, \bibinfo {author} {\bibfnamefont {T.~H.}\ \bibnamefont {Taminiau}}, \emph {et~al.},\ }\bibfield  {title} {\bibinfo {title} {Roadmap on nanoscale magnetic resonance imaging},\ }\href@noop {} {\bibfield  {journal} {\bibinfo  {journal} {Nanotechnology}\ } (\bibinfo {year} {2024})}\BibitemShut {NoStop}%
\bibitem [{\citenamefont {Feng}\ \emph {et~al.}(2022)\citenamefont {Feng}, \citenamefont {Vaghefi}, \citenamefont {Vranjkovic}, \citenamefont {Penedo}, \citenamefont {Kappenberger}, \citenamefont {Schwenk}, \citenamefont {Zhao}, \citenamefont {Mandru},\ and\ \citenamefont {Hug}}]{Feng2022}%
  \BibitemOpen
  \bibfield  {author} {\bibinfo {author} {\bibfnamefont {Y.}~\bibnamefont {Feng}}, \bibinfo {author} {\bibfnamefont {P.~M.}\ \bibnamefont {Vaghefi}}, \bibinfo {author} {\bibfnamefont {S.}~\bibnamefont {Vranjkovic}}, \bibinfo {author} {\bibfnamefont {M.}~\bibnamefont {Penedo}}, \bibinfo {author} {\bibfnamefont {P.}~\bibnamefont {Kappenberger}}, \bibinfo {author} {\bibfnamefont {J.}~\bibnamefont {Schwenk}}, \bibinfo {author} {\bibfnamefont {X.}~\bibnamefont {Zhao}}, \bibinfo {author} {\bibfnamefont {A.-O.}\ \bibnamefont {Mandru}},\ and\ \bibinfo {author} {\bibfnamefont {H.}~\bibnamefont {Hug}},\ }\bibfield  {title} {\bibinfo {title} {{Magnetic force microscopy contrast formation and field sensitivity}},\ }\href {https://doi.org/10.1016/j.jmmm.2022.169073} {\bibfield  {journal} {\bibinfo  {journal} {Journal of Magnetism and Magnetic Materials}\ }\textbf {\bibinfo {volume} {551}},\ \bibinfo {pages} {169073} (\bibinfo {year} {2022})}\BibitemShut {NoStop}%
\bibitem [{\citenamefont {Jaafar}\ \emph {et~al.}(2011)\citenamefont {Jaafar}, \citenamefont {Iglesias-Freire}, \citenamefont {Serrano-Ramon}, \citenamefont {Ibarra}, \citenamefont {Teresa},\ and\ \citenamefont {Asenjo}}]{Jaafar2011}%
  \BibitemOpen
  \bibfield  {author} {\bibinfo {author} {\bibfnamefont {M.}~\bibnamefont {Jaafar}}, \bibinfo {author} {\bibfnamefont {O.}~\bibnamefont {Iglesias-Freire}}, \bibinfo {author} {\bibfnamefont {L.}~\bibnamefont {Serrano-Ramon}}, \bibinfo {author} {\bibfnamefont {M.~R.}\ \bibnamefont {Ibarra}}, \bibinfo {author} {\bibfnamefont {J.~M.~d.}\ \bibnamefont {Teresa}},\ and\ \bibinfo {author} {\bibfnamefont {A.}~\bibnamefont {Asenjo}},\ }\bibfield  {title} {\bibinfo {title} {Distinguishing magnetic and electrostatic interactions by a kelvin probe force microscopy-magnetic force microscopy combination},\ }\href {https://doi.org/10.3762/bjnano.2.59} {\bibfield  {journal} {\bibinfo  {journal} {Beilstein Journal of Nanotechnology}\ }\textbf {\bibinfo {volume} {2}},\ \bibinfo {pages} {552 } (\bibinfo {year} {2011})}\BibitemShut {NoStop}%
\bibitem [{\citenamefont {Tao}\ \emph {et~al.}(2016)\citenamefont {Tao}, \citenamefont {Eichler}, \citenamefont {Holzherr},\ and\ \citenamefont {Degen}}]{Tao2016}%
  \BibitemOpen
  \bibfield  {author} {\bibinfo {author} {\bibfnamefont {Y.}~\bibnamefont {Tao}}, \bibinfo {author} {\bibfnamefont {A.}~\bibnamefont {Eichler}}, \bibinfo {author} {\bibfnamefont {T.}~\bibnamefont {Holzherr}},\ and\ \bibinfo {author} {\bibfnamefont {C.~L.}\ \bibnamefont {Degen}},\ }\bibfield  {title} {\bibinfo {title} {Ultrasensitive mechanical detection of magnetic moment using a commercial disk drive write head},\ }\href {https://doi.org/10.1038/ncomms12714} {\bibfield  {journal} {\bibinfo  {journal} {Nature Communications}\ }\textbf {\bibinfo {volume} {7}},\ \bibinfo {pages} {12714} (\bibinfo {year} {2016})}\BibitemShut {NoStop}%
\bibitem [{\citenamefont {Turchette}\ \emph {et~al.}(2000)\citenamefont {Turchette}, \citenamefont {{Kielpinski}}, \citenamefont {King}, \citenamefont {Leibfried}, \citenamefont {Meekhof}, \citenamefont {Myatt}, \citenamefont {Rowe}, \citenamefont {Sackett}, \citenamefont {Wood}, \citenamefont {Itano}, \citenamefont {Monroe},\ and\ \citenamefont {Wineland}}]{turchette_heating_2000}%
  \BibitemOpen
  \bibfield  {author} {\bibinfo {author} {\bibfnamefont {Q.~A.}\ \bibnamefont {Turchette}}, \bibinfo {author} {\bibnamefont {{Kielpinski}}}, \bibinfo {author} {\bibfnamefont {B.~E.}\ \bibnamefont {King}}, \bibinfo {author} {\bibfnamefont {D.}~\bibnamefont {Leibfried}}, \bibinfo {author} {\bibfnamefont {D.~M.}\ \bibnamefont {Meekhof}}, \bibinfo {author} {\bibfnamefont {C.~J.}\ \bibnamefont {Myatt}}, \bibinfo {author} {\bibfnamefont {M.~A.}\ \bibnamefont {Rowe}}, \bibinfo {author} {\bibfnamefont {C.~A.}\ \bibnamefont {Sackett}}, \bibinfo {author} {\bibfnamefont {C.~S.}\ \bibnamefont {Wood}}, \bibinfo {author} {\bibfnamefont {W.~M.}\ \bibnamefont {Itano}}, \bibinfo {author} {\bibfnamefont {C.}~\bibnamefont {Monroe}},\ and\ \bibinfo {author} {\bibfnamefont {D.~J.}\ \bibnamefont {Wineland}},\ }\bibfield  {title} {\bibinfo {title} {Heating of trapped ions from the quantum ground state},\ }\href {https://doi.org/10.1103/PhysRevA.61.063418} {\bibfield  {journal} {\bibinfo  {journal} {Physical Review A}\ }\textbf
  {\bibinfo {volume} {61}},\ \bibinfo {pages} {063418} (\bibinfo {year} {2000})},\ \bibinfo {note} {publisher: American Physical Society}\BibitemShut {NoStop}%
\bibitem [{\citenamefont {Labaziewicz}\ \emph {et~al.}(2008)\citenamefont {Labaziewicz}, \citenamefont {Ge}, \citenamefont {Leibrandt}, \citenamefont {Wang}, \citenamefont {Shewmon},\ and\ \citenamefont {Chuang}}]{labaziewicz_temperature_2008}%
  \BibitemOpen
  \bibfield  {author} {\bibinfo {author} {\bibfnamefont {J.}~\bibnamefont {Labaziewicz}}, \bibinfo {author} {\bibfnamefont {Y.}~\bibnamefont {Ge}}, \bibinfo {author} {\bibfnamefont {D.~R.}\ \bibnamefont {Leibrandt}}, \bibinfo {author} {\bibfnamefont {S.~X.}\ \bibnamefont {Wang}}, \bibinfo {author} {\bibfnamefont {R.}~\bibnamefont {Shewmon}},\ and\ \bibinfo {author} {\bibfnamefont {I.~L.}\ \bibnamefont {Chuang}},\ }\bibfield  {title} {\bibinfo {title} {Temperature {Dependence} of {Electric} {Field} {Noise} above {Gold} {Surfaces}},\ }\href {https://doi.org/10.1103/PhysRevLett.101.180602} {\bibfield  {journal} {\bibinfo  {journal} {Physical Review Letters}\ }\textbf {\bibinfo {volume} {101}},\ \bibinfo {pages} {180602} (\bibinfo {year} {2008})}\BibitemShut {NoStop}%
\bibitem [{\citenamefont {Safavi-Naini}\ \emph {et~al.}(2011)\citenamefont {Safavi-Naini}, \citenamefont {Rabl}, \citenamefont {Weck},\ and\ \citenamefont {Sadeghpour}}]{safavi-naini_microscopic_2011}%
  \BibitemOpen
  \bibfield  {author} {\bibinfo {author} {\bibfnamefont {A.}~\bibnamefont {Safavi-Naini}}, \bibinfo {author} {\bibfnamefont {P.}~\bibnamefont {Rabl}}, \bibinfo {author} {\bibfnamefont {P.~F.}\ \bibnamefont {Weck}},\ and\ \bibinfo {author} {\bibfnamefont {H.~R.}\ \bibnamefont {Sadeghpour}},\ }\bibfield  {title} {\bibinfo {title} {Microscopic model of electric-field-noise heating in ion traps},\ }\href {https://doi.org/10.1103/PhysRevA.84.023412} {\bibfield  {journal} {\bibinfo  {journal} {Physical Review A}\ }\textbf {\bibinfo {volume} {84}},\ \bibinfo {pages} {023412} (\bibinfo {year} {2011})},\ \bibinfo {note} {publisher: American Physical Society}\BibitemShut {NoStop}%
\bibitem [{\citenamefont {Brownnutt}\ \emph {et~al.}(2015)\citenamefont {Brownnutt}, \citenamefont {Kumph}, \citenamefont {Rabl},\ and\ \citenamefont {Blatt}}]{brownnutt_ion-trap_2015}%
  \BibitemOpen
  \bibfield  {author} {\bibinfo {author} {\bibfnamefont {M.}~\bibnamefont {Brownnutt}}, \bibinfo {author} {\bibfnamefont {M.}~\bibnamefont {Kumph}}, \bibinfo {author} {\bibfnamefont {P.}~\bibnamefont {Rabl}},\ and\ \bibinfo {author} {\bibfnamefont {R.}~\bibnamefont {Blatt}},\ }\bibfield  {title} {\bibinfo {title} {Ion-trap measurements of electric-field noise near surfaces},\ }\href {https://doi.org/10.1103/RevModPhys.87.1419} {\bibfield  {journal} {\bibinfo  {journal} {Reviews of Modern Physics}\ }\textbf {\bibinfo {volume} {87}},\ \bibinfo {pages} {1419} (\bibinfo {year} {2015})},\ \bibinfo {note} {publisher: American Physical Society}\BibitemShut {NoStop}%
\bibitem [{\citenamefont {Kumph}\ \emph {et~al.}(2016)\citenamefont {Kumph}, \citenamefont {Henkel}, \citenamefont {Rabl}, \citenamefont {Brownnutt},\ and\ \citenamefont {Blatt}}]{kumph_electric-field_2016}%
  \BibitemOpen
  \bibfield  {author} {\bibinfo {author} {\bibfnamefont {M.}~\bibnamefont {Kumph}}, \bibinfo {author} {\bibfnamefont {C.}~\bibnamefont {Henkel}}, \bibinfo {author} {\bibfnamefont {P.}~\bibnamefont {Rabl}}, \bibinfo {author} {\bibfnamefont {M.}~\bibnamefont {Brownnutt}},\ and\ \bibinfo {author} {\bibfnamefont {R.}~\bibnamefont {Blatt}},\ }\bibfield  {title} {\bibinfo {title} {Electric-field noise above a thin dielectric layer on metal electrodes},\ }\href {https://doi.org/10.1088/1367-2630/18/2/023020} {\bibfield  {journal} {\bibinfo  {journal} {New Journal of Physics}\ }\textbf {\bibinfo {volume} {18}},\ \bibinfo {pages} {023020} (\bibinfo {year} {2016})},\ \bibinfo {note} {publisher: IOP Publishing}\BibitemShut {NoStop}%
\bibitem [{\citenamefont {Gao}\ \emph {et~al.}(2008)\citenamefont {Gao}, \citenamefont {Daal}, \citenamefont {Vayonakis}, \citenamefont {Kumar}, \citenamefont {Zmuidzinas}, \citenamefont {Sadoulet}, \citenamefont {Mazin}, \citenamefont {Day},\ and\ \citenamefont {Leduc}}]{gao_experimental_2008}%
  \BibitemOpen
  \bibfield  {author} {\bibinfo {author} {\bibfnamefont {J.}~\bibnamefont {Gao}}, \bibinfo {author} {\bibfnamefont {M.}~\bibnamefont {Daal}}, \bibinfo {author} {\bibfnamefont {A.}~\bibnamefont {Vayonakis}}, \bibinfo {author} {\bibfnamefont {S.}~\bibnamefont {Kumar}}, \bibinfo {author} {\bibfnamefont {J.}~\bibnamefont {Zmuidzinas}}, \bibinfo {author} {\bibfnamefont {B.}~\bibnamefont {Sadoulet}}, \bibinfo {author} {\bibfnamefont {B.~A.}\ \bibnamefont {Mazin}}, \bibinfo {author} {\bibfnamefont {P.~K.}\ \bibnamefont {Day}},\ and\ \bibinfo {author} {\bibfnamefont {H.~G.}\ \bibnamefont {Leduc}},\ }\bibfield  {title} {\bibinfo {title} {Experimental evidence for a surface distribution of two-level systems in superconducting lithographed microwave resonators},\ }\href {https://doi.org/10.1063/1.2906373} {\bibfield  {journal} {\bibinfo  {journal} {Applied Physics Letters}\ }\textbf {\bibinfo {volume} {92}},\ \bibinfo {pages} {152505} (\bibinfo {year} {2008})}\BibitemShut {NoStop}%
\bibitem [{\citenamefont {Wang}\ \emph {et~al.}(2009)\citenamefont {Wang}, \citenamefont {Hofheinz}, \citenamefont {Wenner}, \citenamefont {Ansmann}, \citenamefont {Bialczak}, \citenamefont {Lenander}, \citenamefont {Lucero}, \citenamefont {Neeley}, \citenamefont {O’Connell}, \citenamefont {Sank}, \citenamefont {Weides}, \citenamefont {Cleland},\ and\ \citenamefont {Martinis}}]{wang_improving_2009}%
  \BibitemOpen
  \bibfield  {author} {\bibinfo {author} {\bibfnamefont {H.}~\bibnamefont {Wang}}, \bibinfo {author} {\bibfnamefont {M.}~\bibnamefont {Hofheinz}}, \bibinfo {author} {\bibfnamefont {J.}~\bibnamefont {Wenner}}, \bibinfo {author} {\bibfnamefont {M.}~\bibnamefont {Ansmann}}, \bibinfo {author} {\bibfnamefont {R.~C.}\ \bibnamefont {Bialczak}}, \bibinfo {author} {\bibfnamefont {M.}~\bibnamefont {Lenander}}, \bibinfo {author} {\bibfnamefont {E.}~\bibnamefont {Lucero}}, \bibinfo {author} {\bibfnamefont {M.}~\bibnamefont {Neeley}}, \bibinfo {author} {\bibfnamefont {A.~D.}\ \bibnamefont {O’Connell}}, \bibinfo {author} {\bibfnamefont {D.}~\bibnamefont {Sank}}, \bibinfo {author} {\bibfnamefont {M.}~\bibnamefont {Weides}}, \bibinfo {author} {\bibfnamefont {A.~N.}\ \bibnamefont {Cleland}},\ and\ \bibinfo {author} {\bibfnamefont {J.~M.}\ \bibnamefont {Martinis}},\ }\bibfield  {title} {\bibinfo {title} {Improving the coherence time of superconducting coplanar resonators},\ }\href {https://doi.org/10.1063/1.3273372}
  {\bibfield  {journal} {\bibinfo  {journal} {Applied Physics Letters}\ }\textbf {\bibinfo {volume} {95}},\ \bibinfo {pages} {233508} (\bibinfo {year} {2009})}\BibitemShut {NoStop}%
\bibitem [{\citenamefont {Stipe}\ \emph {et~al.}(2001)\citenamefont {Stipe}, \citenamefont {Mamin}, \citenamefont {Stowe}, \citenamefont {Kenny},\ and\ \citenamefont {Rugar}}]{stipe_noncontact_2001}%
  \BibitemOpen
  \bibfield  {author} {\bibinfo {author} {\bibfnamefont {B.~C.}\ \bibnamefont {Stipe}}, \bibinfo {author} {\bibfnamefont {H.~J.}\ \bibnamefont {Mamin}}, \bibinfo {author} {\bibfnamefont {T.~D.}\ \bibnamefont {Stowe}}, \bibinfo {author} {\bibfnamefont {T.~W.}\ \bibnamefont {Kenny}},\ and\ \bibinfo {author} {\bibfnamefont {D.}~\bibnamefont {Rugar}},\ }\bibfield  {title} {\bibinfo {title} {Noncontact {Friction} and {Force} {Fluctuations} between {Closely} {Spaced} {Bodies}},\ }\href {https://doi.org/10.1103/PhysRevLett.87.096801} {\bibfield  {journal} {\bibinfo  {journal} {Physical Review Letters}\ }\textbf {\bibinfo {volume} {87}},\ \bibinfo {pages} {096801} (\bibinfo {year} {2001})}\BibitemShut {NoStop}%
\bibitem [{\citenamefont {Zurita-Sánchez}\ \emph {et~al.}(2004)\citenamefont {Zurita-Sánchez}, \citenamefont {Greffet},\ and\ \citenamefont {Novotny}}]{zurita-sanchez_friction_2004}%
  \BibitemOpen
  \bibfield  {author} {\bibinfo {author} {\bibfnamefont {J.~R.}\ \bibnamefont {Zurita-Sánchez}}, \bibinfo {author} {\bibfnamefont {J.-J.}\ \bibnamefont {Greffet}},\ and\ \bibinfo {author} {\bibfnamefont {L.}~\bibnamefont {Novotny}},\ }\bibfield  {title} {\bibinfo {title} {Friction forces arising from fluctuating thermal fields},\ }\href {https://doi.org/10.1103/PhysRevA.69.022902} {\bibfield  {journal} {\bibinfo  {journal} {Physical Review A}\ }\textbf {\bibinfo {volume} {69}},\ \bibinfo {pages} {022902} (\bibinfo {year} {2004})}\BibitemShut {NoStop}%
\bibitem [{\citenamefont {Kuehn}\ \emph {et~al.}(2006)\citenamefont {Kuehn}, \citenamefont {Loring},\ and\ \citenamefont {Marohn}}]{kuehn_dielectric_2006}%
  \BibitemOpen
  \bibfield  {author} {\bibinfo {author} {\bibfnamefont {S.}~\bibnamefont {Kuehn}}, \bibinfo {author} {\bibfnamefont {R.~F.}\ \bibnamefont {Loring}},\ and\ \bibinfo {author} {\bibfnamefont {J.~A.}\ \bibnamefont {Marohn}},\ }\bibfield  {title} {\bibinfo {title} {Dielectric {Fluctuations} and the {Origins} of {Noncontact} {Friction}},\ }\href {https://doi.org/10.1103/PhysRevLett.96.156103} {\bibfield  {journal} {\bibinfo  {journal} {Physical Review Letters}\ }\textbf {\bibinfo {volume} {96}},\ \bibinfo {pages} {156103} (\bibinfo {year} {2006})}\BibitemShut {NoStop}%
\bibitem [{\citenamefont {Yazdanian}\ \emph {et~al.}(2008)\citenamefont {Yazdanian}, \citenamefont {Marohn},\ and\ \citenamefont {Loring}}]{yazdanian_dielectric_2008}%
  \BibitemOpen
  \bibfield  {author} {\bibinfo {author} {\bibfnamefont {S.~M.}\ \bibnamefont {Yazdanian}}, \bibinfo {author} {\bibfnamefont {J.~A.}\ \bibnamefont {Marohn}},\ and\ \bibinfo {author} {\bibfnamefont {R.~F.}\ \bibnamefont {Loring}},\ }\bibfield  {title} {\bibinfo {title} {Dielectric fluctuations in force microscopy: {Noncontact} friction and frequency jitter},\ }\href {https://doi.org/10.1063/1.2932254} {\bibfield  {journal} {\bibinfo  {journal} {The Journal of Chemical Physics}\ }\textbf {\bibinfo {volume} {128}},\ \bibinfo {pages} {224706} (\bibinfo {year} {2008})}\BibitemShut {NoStop}%
\bibitem [{\citenamefont {Kisiel}\ \emph {et~al.}(2011)\citenamefont {Kisiel}, \citenamefont {Gnecco}, \citenamefont {Gysin}, \citenamefont {Marot}, \citenamefont {Rast},\ and\ \citenamefont {Meyer}}]{kisiel_suppression_2011}%
  \BibitemOpen
  \bibfield  {author} {\bibinfo {author} {\bibfnamefont {M.}~\bibnamefont {Kisiel}}, \bibinfo {author} {\bibfnamefont {E.}~\bibnamefont {Gnecco}}, \bibinfo {author} {\bibfnamefont {U.}~\bibnamefont {Gysin}}, \bibinfo {author} {\bibfnamefont {L.}~\bibnamefont {Marot}}, \bibinfo {author} {\bibfnamefont {S.}~\bibnamefont {Rast}},\ and\ \bibinfo {author} {\bibfnamefont {E.}~\bibnamefont {Meyer}},\ }\bibfield  {title} {\bibinfo {title} {Suppression of electronic friction on {Nb} films in the superconducting state},\ }\href {https://doi.org/10.1038/nmat2936} {\bibfield  {journal} {\bibinfo  {journal} {Nature Materials}\ }\textbf {\bibinfo {volume} {10}},\ \bibinfo {pages} {119} (\bibinfo {year} {2011})}\BibitemShut {NoStop}%
\bibitem [{\citenamefont {She}\ and\ \citenamefont {Balatsky}(2012)}]{she_noncontact_2012}%
  \BibitemOpen
  \bibfield  {author} {\bibinfo {author} {\bibfnamefont {J.-H.}\ \bibnamefont {She}}\ and\ \bibinfo {author} {\bibfnamefont {A.~V.}\ \bibnamefont {Balatsky}},\ }\bibfield  {title} {\bibinfo {title} {Noncontact {Friction} and {Relaxational} {Dynamics} of {Surface} {Defects}},\ }\href {https://doi.org/10.1103/PhysRevLett.108.136101} {\bibfield  {journal} {\bibinfo  {journal} {Physical Review Letters}\ }\textbf {\bibinfo {volume} {108}},\ \bibinfo {pages} {136101} (\bibinfo {year} {2012})}\BibitemShut {NoStop}%
\bibitem [{\citenamefont {den Haan}\ \emph {et~al.}(2015)\citenamefont {den Haan}, \citenamefont {Wagenaar}, \citenamefont {de~Voogd}, \citenamefont {Koning},\ and\ \citenamefont {Oosterkamp}}]{den_haan_spin-mediated_2015}%
  \BibitemOpen
  \bibfield  {author} {\bibinfo {author} {\bibfnamefont {A.~M.~J.}\ \bibnamefont {den Haan}}, \bibinfo {author} {\bibfnamefont {J.~J.~T.}\ \bibnamefont {Wagenaar}}, \bibinfo {author} {\bibfnamefont {J.~M.}\ \bibnamefont {de~Voogd}}, \bibinfo {author} {\bibfnamefont {G.}~\bibnamefont {Koning}},\ and\ \bibinfo {author} {\bibfnamefont {T.~H.}\ \bibnamefont {Oosterkamp}},\ }\bibfield  {title} {\bibinfo {title} {Spin-mediated dissipation and frequency shifts of a cantilever at {milliKelvin} temperatures},\ }\href {https://doi.org/10.1103/PhysRevB.92.235441} {\bibfield  {journal} {\bibinfo  {journal} {Physical Review B}\ }\textbf {\bibinfo {volume} {92}},\ \bibinfo {pages} {235441} (\bibinfo {year} {2015})}\BibitemShut {NoStop}%
\bibitem [{\citenamefont {de~Voogd}\ \emph {et~al.}(2017)\citenamefont {de~Voogd}, \citenamefont {Wagenaar},\ and\ \citenamefont {Oosterkamp}}]{de_voogd_dissipation_2017}%
  \BibitemOpen
  \bibfield  {author} {\bibinfo {author} {\bibfnamefont {J.~M.}\ \bibnamefont {de~Voogd}}, \bibinfo {author} {\bibfnamefont {J.~J.~T.}\ \bibnamefont {Wagenaar}},\ and\ \bibinfo {author} {\bibfnamefont {T.~H.}\ \bibnamefont {Oosterkamp}},\ }\bibfield  {title} {\bibinfo {title} {Dissipation and resonance frequency shift of a resonator magnetically coupled to a semiclassical spin},\ }\href {https://doi.org/10.1038/srep42239} {\bibfield  {journal} {\bibinfo  {journal} {Scientific Reports}\ }\textbf {\bibinfo {volume} {7}},\ \bibinfo {pages} {42239} (\bibinfo {year} {2017})}\BibitemShut {NoStop}%
\bibitem [{\citenamefont {H{\'{e}}ritier}\ \emph {et~al.}(2021)\citenamefont {H{\'{e}}ritier}, \citenamefont {Pachlatko}, \citenamefont {Tao}, \citenamefont {Abendroth}, \citenamefont {Degen},\ and\ \citenamefont {Eichler}}]{heritier_2021}%
  \BibitemOpen
  \bibfield  {author} {\bibinfo {author} {\bibfnamefont {M.}~\bibnamefont {H{\'{e}}ritier}}, \bibinfo {author} {\bibfnamefont {R.}~\bibnamefont {Pachlatko}}, \bibinfo {author} {\bibfnamefont {Y.}~\bibnamefont {Tao}}, \bibinfo {author} {\bibfnamefont {J.~M.}\ \bibnamefont {Abendroth}}, \bibinfo {author} {\bibfnamefont {C.~L.}\ \bibnamefont {Degen}},\ and\ \bibinfo {author} {\bibfnamefont {A.}~\bibnamefont {Eichler}},\ }\bibfield  {title} {\bibinfo {title} {Spatial correlation between fluctuating and static fields over metal and dielectric substrates},\ }\bibfield  {journal} {\bibinfo  {journal} {Physical Review Letters}\ }\textbf {\bibinfo {volume} {127}},\ \href {https://doi.org/10.1103/physrevlett.127.216101} {10.1103/physrevlett.127.216101} (\bibinfo {year} {2021})\BibitemShut {NoStop}%
\bibitem [{\citenamefont {Rosskopf}\ \emph {et~al.}(2014)\citenamefont {Rosskopf}, \citenamefont {Dussaux}, \citenamefont {Ohashi}, \citenamefont {Loretz}, \citenamefont {Schirhagl}, \citenamefont {Watanabe}, \citenamefont {Shikata}, \citenamefont {Itoh},\ and\ \citenamefont {Degen}}]{PhysRevLett.112.147602}%
  \BibitemOpen
  \bibfield  {author} {\bibinfo {author} {\bibfnamefont {T.}~\bibnamefont {Rosskopf}}, \bibinfo {author} {\bibfnamefont {A.}~\bibnamefont {Dussaux}}, \bibinfo {author} {\bibfnamefont {K.}~\bibnamefont {Ohashi}}, \bibinfo {author} {\bibfnamefont {M.}~\bibnamefont {Loretz}}, \bibinfo {author} {\bibfnamefont {R.}~\bibnamefont {Schirhagl}}, \bibinfo {author} {\bibfnamefont {H.}~\bibnamefont {Watanabe}}, \bibinfo {author} {\bibfnamefont {S.}~\bibnamefont {Shikata}}, \bibinfo {author} {\bibfnamefont {K.~M.}\ \bibnamefont {Itoh}},\ and\ \bibinfo {author} {\bibfnamefont {C.~L.}\ \bibnamefont {Degen}},\ }\bibfield  {title} {\bibinfo {title} {Investigation of surface magnetic noise by shallow spins in diamond},\ }\href {https://doi.org/10.1103/PhysRevLett.112.147602} {\bibfield  {journal} {\bibinfo  {journal} {Phys. Rev. Lett.}\ }\textbf {\bibinfo {volume} {112}},\ \bibinfo {pages} {147602} (\bibinfo {year} {2014})}\BibitemShut {NoStop}%
\bibitem [{\citenamefont {Ko\ifmmode~\check{s}\else \v{s}\fi{}ata}\ \emph {et~al.}(2020)\citenamefont {Ko\ifmmode~\check{s}\else \v{s}\fi{}ata}, \citenamefont {Zilberberg}, \citenamefont {Degen}, \citenamefont {Chitra},\ and\ \citenamefont {Eichler}}]{Kosata_2020}%
  \BibitemOpen
  \bibfield  {author} {\bibinfo {author} {\bibfnamefont {J.}~\bibnamefont {Ko\ifmmode~\check{s}\else \v{s}\fi{}ata}}, \bibinfo {author} {\bibfnamefont {O.}~\bibnamefont {Zilberberg}}, \bibinfo {author} {\bibfnamefont {C.~L.}\ \bibnamefont {Degen}}, \bibinfo {author} {\bibfnamefont {R.}~\bibnamefont {Chitra}},\ and\ \bibinfo {author} {\bibfnamefont {A.}~\bibnamefont {Eichler}},\ }\bibfield  {title} {\bibinfo {title} {Spin detection via parametric frequency conversion in a membrane resonator},\ }\href {https://doi.org/10.1103/PhysRevApplied.14.014042} {\bibfield  {journal} {\bibinfo  {journal} {Phys. Rev. Applied}\ }\textbf {\bibinfo {volume} {14}},\ \bibinfo {pages} {014042} (\bibinfo {year} {2020})}\BibitemShut {NoStop}%
\bibitem [{\citenamefont {H\"alg}\ \emph {et~al.}(2021{\natexlab{a}})\citenamefont {H\"alg}, \citenamefont {Gisler}, \citenamefont {Tsaturyan}, \citenamefont {Catalini}, \citenamefont {Grob}, \citenamefont {Krass}, \citenamefont {H\'eritier}, \citenamefont {Mattiat}, \citenamefont {Thamm}, \citenamefont {Schirhagl}, \citenamefont {Langman}, \citenamefont {Schliesser}, \citenamefont {Degen},\ and\ \citenamefont {Eichler}}]{Halg_2022}%
  \BibitemOpen
  \bibfield  {author} {\bibinfo {author} {\bibfnamefont {D.}~\bibnamefont {H\"alg}}, \bibinfo {author} {\bibfnamefont {T.}~\bibnamefont {Gisler}}, \bibinfo {author} {\bibfnamefont {Y.}~\bibnamefont {Tsaturyan}}, \bibinfo {author} {\bibfnamefont {L.}~\bibnamefont {Catalini}}, \bibinfo {author} {\bibfnamefont {U.}~\bibnamefont {Grob}}, \bibinfo {author} {\bibfnamefont {M.-D.}\ \bibnamefont {Krass}}, \bibinfo {author} {\bibfnamefont {M.}~\bibnamefont {H\'eritier}}, \bibinfo {author} {\bibfnamefont {H.}~\bibnamefont {Mattiat}}, \bibinfo {author} {\bibfnamefont {A.-K.}\ \bibnamefont {Thamm}}, \bibinfo {author} {\bibfnamefont {R.}~\bibnamefont {Schirhagl}}, \bibinfo {author} {\bibfnamefont {E.~C.}\ \bibnamefont {Langman}}, \bibinfo {author} {\bibfnamefont {A.}~\bibnamefont {Schliesser}}, \bibinfo {author} {\bibfnamefont {C.~L.}\ \bibnamefont {Degen}},\ and\ \bibinfo {author} {\bibfnamefont {A.}~\bibnamefont {Eichler}},\ }\bibfield  {title} {\bibinfo {title} {Membrane-based scanning force microscopy},\ }\href
  {https://doi.org/10.1103/PhysRevApplied.15.L021001} {\bibfield  {journal} {\bibinfo  {journal} {Phys. Rev. Appl.}\ }\textbf {\bibinfo {volume} {15}},\ \bibinfo {pages} {L021001} (\bibinfo {year} {2021}{\natexlab{a}})}\BibitemShut {NoStop}%
\bibitem [{\citenamefont {Longenecker}\ \emph {et~al.}(2012)\citenamefont {Longenecker}, \citenamefont {Mamin}, \citenamefont {Senko}, \citenamefont {Chen}, \citenamefont {Rettner}, \citenamefont {Rugar},\ and\ \citenamefont {Marohn}}]{longenecker2012high}%
  \BibitemOpen
  \bibfield  {author} {\bibinfo {author} {\bibfnamefont {J.~G.}\ \bibnamefont {Longenecker}}, \bibinfo {author} {\bibfnamefont {H.}~\bibnamefont {Mamin}}, \bibinfo {author} {\bibfnamefont {A.~W.}\ \bibnamefont {Senko}}, \bibinfo {author} {\bibfnamefont {L.}~\bibnamefont {Chen}}, \bibinfo {author} {\bibfnamefont {C.~T.}\ \bibnamefont {Rettner}}, \bibinfo {author} {\bibfnamefont {D.}~\bibnamefont {Rugar}},\ and\ \bibinfo {author} {\bibfnamefont {J.~A.}\ \bibnamefont {Marohn}},\ }\bibfield  {title} {\bibinfo {title} {High-gradient nanomagnets on cantilevers for sensitive detection of nuclear magnetic resonance},\ }\href {https://doi.org/10.1021/nn3030628} {\bibfield  {journal} {\bibinfo  {journal} {ACS nano}\ }\textbf {\bibinfo {volume} {6}},\ \bibinfo {pages} {9637} (\bibinfo {year} {2012})},\ \bibinfo {note} {publisher: ACS Publications}\BibitemShut {NoStop}%
\bibitem [{\citenamefont {H\"alg}\ \emph {et~al.}(2021{\natexlab{b}})\citenamefont {H\"alg}, \citenamefont {Gisler}, \citenamefont {Tsaturyan}, \citenamefont {Catalini}, \citenamefont {Grob}, \citenamefont {Krass}, \citenamefont {H\'eritier}, \citenamefont {Mattiat}, \citenamefont {Thamm}, \citenamefont {Schirhagl}, \citenamefont {Langman}, \citenamefont {Schliesser}, \citenamefont {Degen},\ and\ \citenamefont {Eichler}}]{Halg_2021}%
  \BibitemOpen
  \bibfield  {author} {\bibinfo {author} {\bibfnamefont {D.}~\bibnamefont {H\"alg}}, \bibinfo {author} {\bibfnamefont {T.}~\bibnamefont {Gisler}}, \bibinfo {author} {\bibfnamefont {Y.}~\bibnamefont {Tsaturyan}}, \bibinfo {author} {\bibfnamefont {L.}~\bibnamefont {Catalini}}, \bibinfo {author} {\bibfnamefont {U.}~\bibnamefont {Grob}}, \bibinfo {author} {\bibfnamefont {M.-D.}\ \bibnamefont {Krass}}, \bibinfo {author} {\bibfnamefont {M.}~\bibnamefont {H\'eritier}}, \bibinfo {author} {\bibfnamefont {H.}~\bibnamefont {Mattiat}}, \bibinfo {author} {\bibfnamefont {A.-K.}\ \bibnamefont {Thamm}}, \bibinfo {author} {\bibfnamefont {R.}~\bibnamefont {Schirhagl}}, \bibinfo {author} {\bibfnamefont {E.~C.}\ \bibnamefont {Langman}}, \bibinfo {author} {\bibfnamefont {A.}~\bibnamefont {Schliesser}}, \bibinfo {author} {\bibfnamefont {C.~L.}\ \bibnamefont {Degen}},\ and\ \bibinfo {author} {\bibfnamefont {A.}~\bibnamefont {Eichler}},\ }\bibfield  {title} {\bibinfo {title} {Membrane-based scanning force microscopy},\ }\href
  {https://doi.org/10.1103/PhysRevApplied.15.L021001} {\bibfield  {journal} {\bibinfo  {journal} {Phys. Rev. Applied}\ }\textbf {\bibinfo {volume} {15}},\ \bibinfo {pages} {L021001} (\bibinfo {year} {2021}{\natexlab{b}})}\BibitemShut {NoStop}%
\bibitem [{\citenamefont {Eichler}(2022)}]{eichler2022ultra}%
  \BibitemOpen
  \bibfield  {author} {\bibinfo {author} {\bibfnamefont {A.}~\bibnamefont {Eichler}},\ }\bibfield  {title} {\bibinfo {title} {Ultra-high-q nanomechanical resonators for force sensing},\ }\href {https://doi.org/10.1088/2633-4356/acaba4} {\bibfield  {journal} {\bibinfo  {journal} {Materials for Quantum Technology}\ }\textbf {\bibinfo {volume} {2}},\ \bibinfo {pages} {043001} (\bibinfo {year} {2022})}\BibitemShut {NoStop}%
\bibitem [{\citenamefont {Bereyhi}\ \emph {et~al.}(2022)\citenamefont {Bereyhi}, \citenamefont {Arabmoheghi}, \citenamefont {Beccari}, \citenamefont {Fedorov}, \citenamefont {Huang}, \citenamefont {Kippenberg},\ and\ \citenamefont {Engelsen}}]{Bereyhi_2022}%
  \BibitemOpen
  \bibfield  {author} {\bibinfo {author} {\bibfnamefont {M.~J.}\ \bibnamefont {Bereyhi}}, \bibinfo {author} {\bibfnamefont {A.}~\bibnamefont {Arabmoheghi}}, \bibinfo {author} {\bibfnamefont {A.}~\bibnamefont {Beccari}}, \bibinfo {author} {\bibfnamefont {S.~A.}\ \bibnamefont {Fedorov}}, \bibinfo {author} {\bibfnamefont {G.}~\bibnamefont {Huang}}, \bibinfo {author} {\bibfnamefont {T.~J.}\ \bibnamefont {Kippenberg}},\ and\ \bibinfo {author} {\bibfnamefont {N.~J.}\ \bibnamefont {Engelsen}},\ }\bibfield  {title} {\bibinfo {title} {Perimeter modes of nanomechanical resonators exhibit quality factors exceeding ${10}^{9}$ at room temperature},\ }\href {https://doi.org/10.1103/PhysRevX.12.021036} {\bibfield  {journal} {\bibinfo  {journal} {Phys. Rev. X}\ }\textbf {\bibinfo {volume} {12}},\ \bibinfo {pages} {021036} (\bibinfo {year} {2022})}\BibitemShut {NoStop}%
\end{thebibliography}%


\providecommand{\noopsort}[1]{}\providecommand{\singleletter}[1]{#1}%
\begin{thebibliography}{2}%
\makeatletter
\providecommand \@ifxundefined [1]{%
 \@ifx{#1\undefined}
}%
\providecommand \@ifnum [1]{%
 \ifnum #1\expandafter \@firstoftwo
 \else \expandafter \@secondoftwo
 \fi
}%
\providecommand \@ifx [1]{%
 \ifx #1\expandafter \@firstoftwo
 \else \expandafter \@secondoftwo
 \fi
}%
\providecommand \natexlab [1]{#1}%
\providecommand \enquote  [1]{``#1''}%
\providecommand \bibnamefont  [1]{#1}%
\providecommand \bibfnamefont [1]{#1}%
\providecommand \citenamefont [1]{#1}%
\providecommand \href@noop [0]{\@secondoftwo}%
\providecommand \href [0]{\begingroup \@sanitize@url \@href}%
\providecommand \@href[1]{\@@startlink{#1}\@@href}%
\providecommand \@@href[1]{\endgroup#1\@@endlink}%
\providecommand \@sanitize@url [0]{\catcode `\\12\catcode `\$12\catcode `\&12\catcode `\#12\catcode `\^12\catcode `\_12\catcode `\%12\relax}%
\providecommand \@@startlink[1]{}%
\providecommand \@@endlink[0]{}%
\providecommand \url  [0]{\begingroup\@sanitize@url \@url }%
\providecommand \@url [1]{\endgroup\@href {#1}{\urlprefix }}%
\providecommand \urlprefix  [0]{URL }%
\providecommand \Eprint [0]{\href }%
\providecommand \doibase [0]{https://doi.org/}%
\providecommand \selectlanguage [0]{\@gobble}%
\providecommand \bibinfo  [0]{\@secondoftwo}%
\providecommand \bibfield  [0]{\@secondoftwo}%
\providecommand \translation [1]{[#1]}%
\providecommand \BibitemOpen [0]{}%
\providecommand \bibitemStop [0]{}%
\providecommand \bibitemNoStop [0]{.\EOS\space}%
\providecommand \EOS [0]{\spacefactor3000\relax}%
\providecommand \BibitemShut  [1]{\csname bibitem#1\endcsname}%
\let\auto@bib@innerbib\@empty
\bibitem [{\citenamefont {Gisler}\ \emph {et~al.}(2024)\citenamefont {Gisler}, \citenamefont {H\"alg}, \citenamefont {Dumont}, \citenamefont {Misra}, \citenamefont {Catalini}, \citenamefont {Langman}, \citenamefont {Schliesser}, \citenamefont {Degen},\ and\ \citenamefont {Eichler}}]{PhysRevApplied.22.044001}%
  \BibitemOpen
  \bibfield  {author} {\bibinfo {author} {\bibfnamefont {T.}~\bibnamefont {Gisler}}, \bibinfo {author} {\bibfnamefont {D.}~\bibnamefont {H\"alg}}, \bibinfo {author} {\bibfnamefont {V.}~\bibnamefont {Dumont}}, \bibinfo {author} {\bibfnamefont {S.}~\bibnamefont {Misra}}, \bibinfo {author} {\bibfnamefont {L.}~\bibnamefont {Catalini}}, \bibinfo {author} {\bibfnamefont {E.~C.}\ \bibnamefont {Langman}}, \bibinfo {author} {\bibfnamefont {A.}~\bibnamefont {Schliesser}}, \bibinfo {author} {\bibfnamefont {C.~L.}\ \bibnamefont {Degen}},\ and\ \bibinfo {author} {\bibfnamefont {A.}~\bibnamefont {Eichler}},\ }\bibfield  {title} {\bibinfo {title} {Enhancing membrane-based scanning force microscopy through an optical cavity},\ }\href {https://doi.org/10.1103/PhysRevApplied.22.044001} {\bibfield  {journal} {\bibinfo  {journal} {Phys. Rev. Appl.}\ }\textbf {\bibinfo {volume} {22}},\ \bibinfo {pages} {044001} (\bibinfo {year} {2024})}\BibitemShut {NoStop}%
\bibitem [{\citenamefont {Feng}\ \emph {et~al.}(2022)\citenamefont {Feng}, \citenamefont {Vaghefi}, \citenamefont {Vranjkovic}, \citenamefont {Penedo}, \citenamefont {Kappenberger}, \citenamefont {Schwenk}, \citenamefont {Zhao}, \citenamefont {Mandru},\ and\ \citenamefont {Hug}}]{Feng2022}%
  \BibitemOpen
  \bibfield  {author} {\bibinfo {author} {\bibfnamefont {Y.}~\bibnamefont {Feng}}, \bibinfo {author} {\bibfnamefont {P.~M.}\ \bibnamefont {Vaghefi}}, \bibinfo {author} {\bibfnamefont {S.}~\bibnamefont {Vranjkovic}}, \bibinfo {author} {\bibfnamefont {M.}~\bibnamefont {Penedo}}, \bibinfo {author} {\bibfnamefont {P.}~\bibnamefont {Kappenberger}}, \bibinfo {author} {\bibfnamefont {J.}~\bibnamefont {Schwenk}}, \bibinfo {author} {\bibfnamefont {X.}~\bibnamefont {Zhao}}, \bibinfo {author} {\bibfnamefont {A.-O.}\ \bibnamefont {Mandru}},\ and\ \bibinfo {author} {\bibfnamefont {H.}~\bibnamefont {Hug}},\ }\bibfield  {title} {\bibinfo {title} {{Magnetic force microscopy contrast formation and field sensitivity}},\ }\href {https://doi.org/10.1016/j.jmmm.2022.169073} {\bibfield  {journal} {\bibinfo  {journal} {Journal of Magnetism and Magnetic Materials}\ }\textbf {\bibinfo {volume} {551}},\ \bibinfo {pages} {169073} (\bibinfo {year} {2022})}\BibitemShut {NoStop}%
\end{thebibliography}%

\end{document}


\title{\textbf{Supplemental Material for:}\\ Differential Magnetic Force Microscopy with a Switchable Tip}
\author{Shobhna Misra}
\affiliation{\affilETH}

\author{Reshma Peremadathil-Pradeep}
\affiliation{\affilEMPA}

\author{Yaoxuan Feng}
\affiliation{\affilEMPA}

\author{Urs Grob}
\affiliation{\affilETH}

\author{Andrada Oana Mandru}
\affiliation{\affilEMPA}

\author{Christian L. Degen}
\affiliation{\affilETH}
\affiliation{\affilETHQuantum}

\author{Hans J. Hug}
\affiliation{\affilEMPA}
\affiliation{\affilUnibas}

\author{Alexander Eichler}
\email[Corresponding author: ]{eichlera@ethz.ch}
\affiliation{\affilETH}
\maketitle

\spacing{1.5}

\section{Numerical simulations of the magnetic tip}
\begin{figure}
\includegraphics[width=\columnwidth]{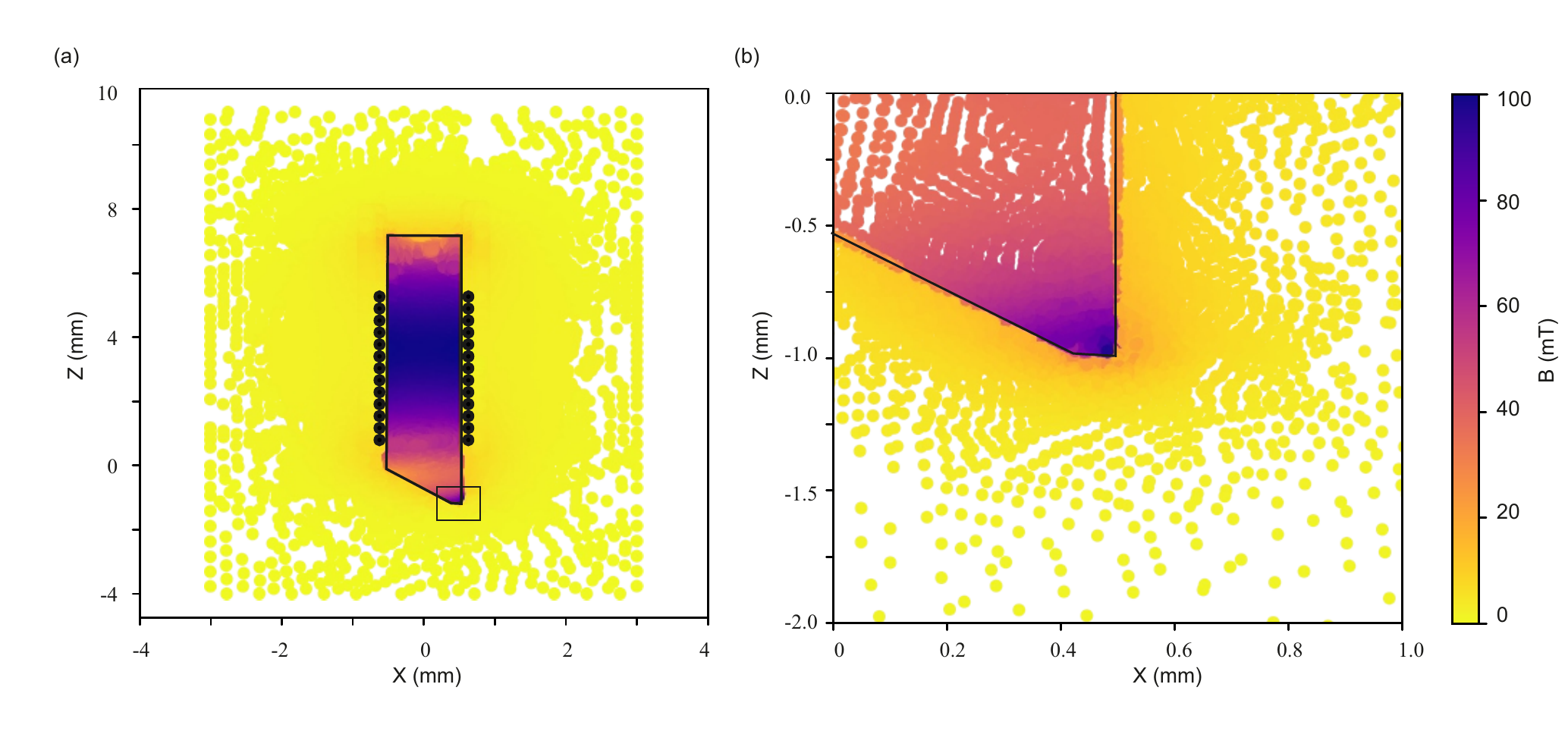} 
\caption{Simulated flux density (a) in the electromagnet and (b) in a cross section close to the tip.
}
\label{fig:sim}
\end{figure}

Our switchable magnetic tip has an open-core design that is compatible with various applications, including the membrane-in-the-middle cavity we plan to employ in the future \cite{PhysRevApplied.22.044001}. This flexibility comes at the cost of a lower magnetic flux density than what can be achieved with a closed magnetic design, as is used in write heads. The target magnetic flux density for fully magnetizing the coated tip is know to be \SI{30}{\milli\tesla}~\cite{Feng2022}. We performed finite element simulations in COMSOL to optimize the design of the ferrite core and of the coil, see Fig.~\ref{fig:sim}. With these simulations, we found that the soft magnetic core (made from MnZn) increases the flux density at the tip location by a factor of 40 compared to an air core.

\section{Tip Fabrication}
The cores for our switchable tips are machined by the machines workshop at the Department of Physics at ETH Zurich. Starting from the bulk ferrite material, the workshop shapes the tapered end of the cores and then fabricates the cylindrical pieces of \SI{1}{\milli\meter} diameter and lengths of roughly \SI{10}{\milli\meter} with a micromill. The finished cores are sent to Sibatron AG, where copper wires of \SI{50}{\micro\meter} diameter are wound around the cores, resulting in an electromagnet. In parallel, commercially available AFM cantilevers (Nanotec SS-ISC) are sputter-coated with Co. Finally, we use a micromanipulator stage under an optical microscope with a large working distance to attach magnetically coated cantilevers to electromagnets. We use the EPOTEC epoxy H20E, which is a conductive two-part glue, to ensure electrical conduction between the ferrite core and the attached cantilever, which is essential for our magnetic force detection protocol. The cantilevers are broken from their support chip using a sharp glass ferrule.

\section{Electronics: Home-built oscillator}

\begin{figure}
\includegraphics[width=\columnwidth]{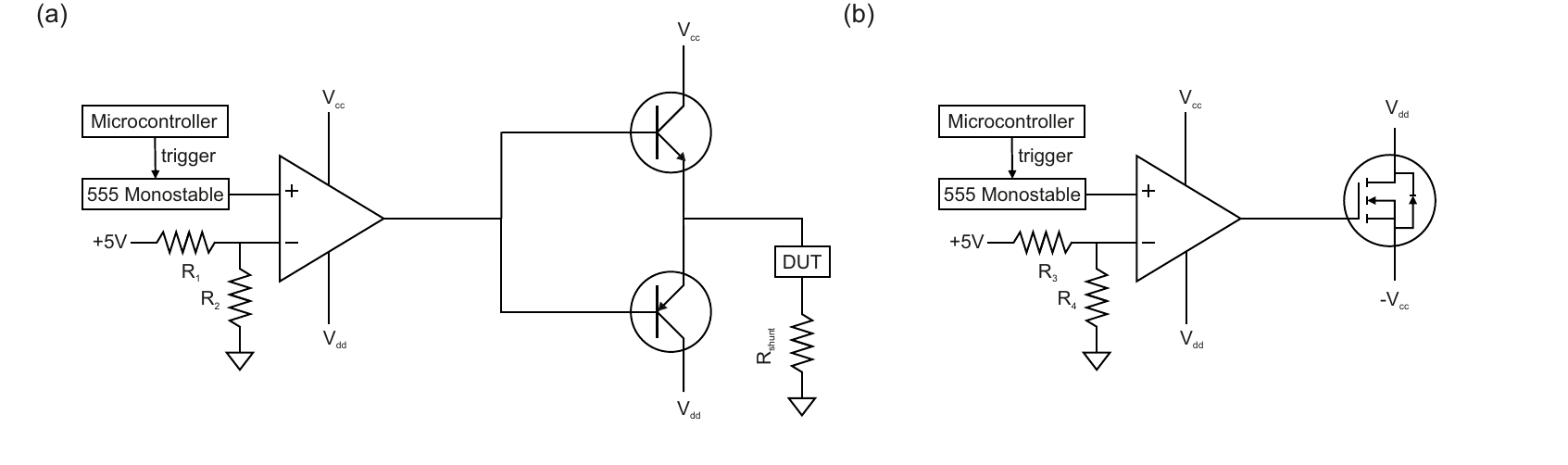} 
\caption{Schematic of the circuit used to generate the current pulses.
}
\label{fig:ckt}
\end{figure}

We use a home-built circuit for generating square pulses with a controlled current amplitude, duration, and duty cycle, see Fig.~\ref{fig:ckt}. The circuit consists of a microcontroller to trigger a 555 timer used in the monostable configuration, which gives out pulses of a precise duration. The pulse duration is controlled by the variable resistor and capacitor in the RC circuit of the 555 monostable block. The output of the timer is fed into a rail-to-rail op-amp comparator that changes the output range from \SI{0}{\volt}-\SI{5}{\volt} to $V_{\rm dd}$-$V_{\rm cc}$, where $V_{\rm dd}$ is either 0V for single pulses or -$V_{\rm cc}$ for a continuous pulse train. The circuit block that controls $V_{\rm dd}$ is shown in \ref{fig:ckt}(b). The output of the comparator is fed into the base of two transistors that operate as switches, which are on only one at a time. The emitter is connected to the device under test (DUT, in our setup usually the coil) and a shunt. The voltage across the shunt is monitored with an oscilloscope to record the current through the DUT.     

\section{Measured current pulses}

\begin{figure}
\includegraphics[width=\columnwidth]{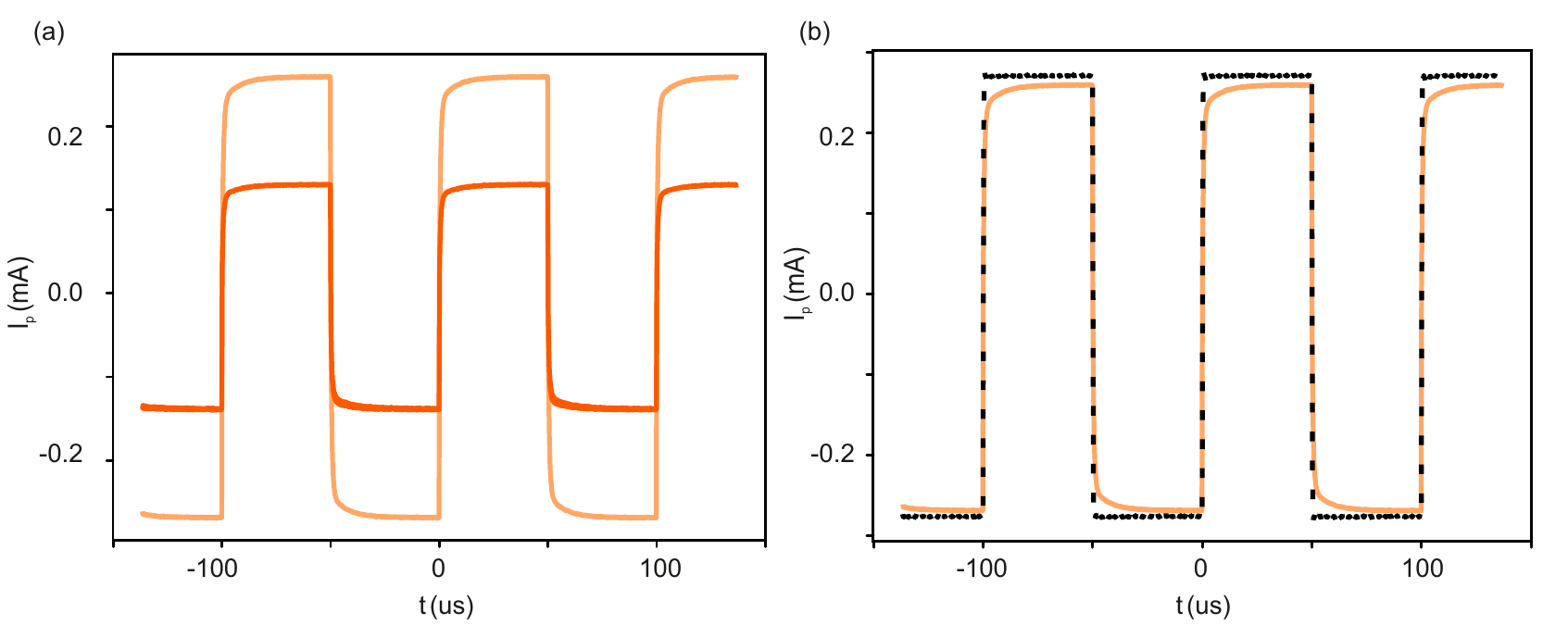} 
\caption{Current pulses (a) through the coil with different amplitudes and (b) through the coil (coral) and through a resistor of similar DC resistance (black).
}
\label{fig:pulse}
\end{figure}

The current measured to flow through the coil in response to a square voltage pulse is shown in Fig.~\ref{fig:pulse}. We observe a rise time of approximately \SI{25}{\micro\second} for the current to saturate, which agrees well with the frequency limit found for reliable tip switching. For a resistor of similar DC resistance as the coil, the response is a step function, as can be seen in Fig.~\ref{fig:pulse}(b).    

\bibliography{references.bib}